\documentclass[sigconf]{acmart}

\copyrightyear{2022}
\acmYear{2022}
\setcopyright{acmlicensed}
\acmConference[WSDM '22] {Proceedings of the Fifteenth ACM International Conference on Web Search and Data Mining}{February 21--25, 2022}{Tempe, AZ, USA.}
\acmBooktitle{Proceedings of the Fifteenth ACM International Conference on Web Search and Data Mining (WSDM '22), February 21--25, 2022, Tempe, AZ, USA}
\acmPrice{15.00}
\acmISBN{978-1-4503-9132-0/22/02}
\acmDOI{10.1145/3488560.3498457}

\AtBeginDocument{%
  \providecommand\BibTeX{{%
    \normalfont B\kern-0.5em{\scshape i\kern-0.25em b}\kern-0.8em\TeX}}}

\usepackage{amsfonts}
\usepackage{algorithmicx}
\usepackage{algorithm}
\usepackage{algpseudocode}
\usepackage{bm}
\usepackage{textcomp}
\usepackage{subfigure}

\usepackage{balance}

\usepackage[utf8]{inputenc}
\def\BibTeX{{\rm B\kern-.05em{\sc i\kern-.025em b}\kern-.08em
    T\kern-.1667em\lower.7ex\hbox{E}\kern-.125emX}}
    
\begin{document}
\fancyhead{}
\title{Identifying Cost-effective Debunkers for Multi-stage Fake News Mitigation Campaigns}

\author{Xiaofei Xu}
\orcid{0000-0001-6988-7541}
\affiliation{%
  \institution{School of Computing Technologies, RMIT University}
  \city{Melbourne}
  \state{Victoria}
  \country{Australia}
}
\email{s3833028@student.rmit.edu.au}

\author{Ke Deng}
\orcid{0000-0002-1008-2498}
\affiliation{%
  \institution{School of Computing Technologies, RMIT University}
  \city{Melbourne}
  \state{Victoria}
  \country{Australia}
}
\email{ke.deng@rmit.edu.au}

\author{Xiuzhen Zhang}
\authornote{Xiuzhen Zhang is the corresponding author.}
\orcid{0000-0001-5558-3790}
\affiliation{%
  \institution{School of Computing Technologies, RMIT University}
  \city{Melbourne}
  \state{Victoria}
  \country{Australia}
}
\email{xiuzhen.zhang@rmit.edu.au}

\begin{abstract}
Online social networks have become a fertile ground for spreading fake news. 
Methods to automatically mitigate fake news propagation have been proposed. Some studies focus on selecting top $k$ influential users on social networks as debunkers, but the social influence of debunkers may not translate to wide mitigation information propagation as expected. 
Other studies assume a given set of debunkers and focus on optimizing intensity for debunkers to publish true news, but as debunkers are fixed, even if with high social influence and/or high intensity to post true news, the true news may not reach users exposed to fake news and therefore mitigation effect may be limited. 
In this paper, we propose the multi-stage fake news mitigation campaign where debunkers are dynamically selected within budget at each stage. We formulate it as a reinforcement learning problem and propose a greedy algorithm optimized by predicting future states so that the debunkers can be selected in a way that maximizes the overall mitigation effect. We conducted extensive experiments on synthetic and real-world social networks and show that our solution outperforms state-of-the-art baselines in terms of mitigation effect. 

\end{abstract}



\maketitle

\section{Introduction}
With the rapid development of the Internet and mobile devices, people tend to spend more time online and interact with others through social network platforms. Although the social network brings real-time and free news, without the professional editing service, the credibility of news on the social network is lower than traditional media sources. For example, one can intentionally generate some misleading news mixed with true news and spread them on the social network \cite{allcott2017social}. Despite acknowledging that there might exist misleading information on social networks, about half (53\%) of U.S. adults say they get news from social networks “often” or “sometimes”.~\footnote{https://www.journalism.org/2021/01/12/news-use-across-social-media-platforms-in-2020/} The spread of fake news at such a scale poses threat to online information security for the online population and society.

To counteract fake news and misinformation, there have been manual fact-checking services as well as algorithms for automatic fake news detection (See~\cite{shu2019studying} for a survey). Notwithstanding these efforts for fake news detection, to effectively counteract fake news at the network scale, arguably it is more important to actively propagate the true news containing correction information such as checked-facts to mitigate the spread of fake news on social networks. 

There have been limited studies on automatic fake news mitigation. Some studies heuristically select users of high social influence -- having a large number of followers -- as debunkers~\cite{saxena2020k, saxena2020mitigating} to propagate true news to mitigate the spread of fake news. The assumption is that influential users on the social network produce wide propagation of true news. But research has shown that overall influence on the social network may not translate to wide mitigation information propagation as expected~\cite{farajtabar2016multistage}. A recent study assumes a given set of users as debunkers who can broadcast true news to mitigate the spread of fake news~\cite{farajtabar2017fake}. A reinforcement learning agent is trained to decide the optimal intensity for debunkers to post true news with the aim to mitigate fake news spread. 
Given the dynamic propagation of fake news on the social network, 
the true news posted by specific debunkers, even if in high intensity, may not reach users exposed to fake news. As a result, this approach may not achieve effective mitigation for the whole social network. In summary, existing studies have not considered how to dynamically select debunkers. 

In this paper, we study the problem of selecting debunkers for the multi-stage fake news mitigation campaign 
and formulate it as a reinforcement learning problem.~\footnote{A mitigation campaign is also known as an episode in reinforcement learning.} Based on the {\em current propagation state} of fake and true news on social networks, our solution dynamically selects debunkers within budget at each stage with the objective to maximize the cumulative mitigation effect -- more true news will be propagated to users exposed to more fake news -- across stages of the campaign. Different from studies by Saxena~\cite{saxena2020k,saxena2020mitigating}, our selected debunkers are not necessarily the most influential users on social networks but only those with maximum influence over the users who have been exposed to fake news. Different from the study by Farajtabar~\cite{farajtabar2017fake}, our debunkers are dynamically selected for each stage according to the fake news propagation state at the time, that is, the set of debunkers may differ from stage to stage. 

%

It is highly challenging to select multiple debunkers within budget at each stage to achieve optimal fake news mitigation. The search space includes all possible user combinations which increase exponentially with the number of users on social networks. So, a greedy strategy can be adopted where a mitigation policy is trained via a reinforcement learning framework to select one user with the highest cumulative mitigation reward as the debunker, and the policy is repeatedly applied to select multiple debunkers until the budget is exhausted. Such greedy strategy however, may have significant mitigation overlap -- the true news posted by debunkers are received by the same users -- and as a result, the overall mitigation effect is limited. To address this issue, we propose a greedy algorithm optimized by predicting future states
so that the debunkers can be selected in a way that minimizes mitigation overlap and maximizes the overall mitigation effect. Our proposed model DQN-FSP extends the deep $Q$-network~\cite{mnih2015human} with future state prediction via the RNN model.

We conducted extensive experiments with synthetic and real-world social network datasets to evaluate DQN-FSP. Experiment results show that DQN-FSP outperforms state-of-the-art baselines. In summary, the contributions of our study are threefold: 
\vspace{-0.1cm}
\begin{itemize}
    \item We introduce the problem of selecting cost-effective debunkers within budget for multi-stage mitigation campaigns.
    \item We formulate the campaign as a reinforcement learning problem to train a mitigation policy that optimizes debunker selection at each stage to maximize the cumulative mitigation across stages for the campaign. 
    \item We propose a greedy algorithm optimized with the RNN model to minimize the mitigation overlap between selected debunkers to improve the overall mitigation effect. 
\end{itemize}

\section{Related Work}\label{related_work}
There have been many studies on the detection of fake news on social network. To reduce the cost and time of manual fact-checking, automatic detection of fake news and prediction of the credibility for social network posts have been proposed, using features such as network features \cite{benamira2019semi}, multi-modal features \cite{wang2018eann} or combined features \cite{shu2017fake}\cite{shu2019studying}. Other studies detect fake news spreaders on social networks using linguistic and personality features \cite{shrestha2020detecting}.

Beyond fake news detection, research on strategies for posting counter true news such as fact-checked contents to mitigate the spreading of fake news on the social network is attracting more attention. Mitigation studies can be categorised into two main classes. One class of studies focus on selecting debunkers to maximize the spread of truthful information to counteract the fake news spread on social networks. Some studies heuristically select top $k$ most influential users as debunkers~\cite{saxena2020k, saxena2020mitigating}. Their assumption is that users with high social influence produce wide propagation of true news on social networks. But research has shown that overall influence on the social network may not translate to wide mitigation information propagation as expected ~\cite{farajtabar2016multistage}.

Another class of studies focus on optimizing the intensity of posting true news for a given set of debunkers to counteract fake news spread on social networks~\cite{farajtabar2017fake, goindani2020social, goindani2020cluster}. As true news are only posted by specific debunkers, even if in high intensity, their propagation may not reach users frequently exposed to fake news given the unknown origin and dynamic propagation of fake news on the social network; the mitigation effect may not be optimal. 
 
Our study falls into the first class of studies of selecting debunkers. But different from previous studies of selecting $k$ debunkers~\cite{saxena2020k, saxena2020mitigating}, we focus on selecting cost-effective debunkers within budget for a multi-stage mitigation campaign. 
Considering the dynamic news diffusion behaviour of users, we aim for a multi-stage mitigation policy such that each stage dynamically selects debunkers according to the {\em current} propagation state of fake news while at the same time achieving maximal mitigation effect across stages.

Reinforcement learning (RL) has been used in fake news mitigation research~\cite{farajtabar2017fake,goindani2020social}. The objective is to predict the intensity for posting true news given specific debunkers, where the value of intensity is continuous. The reinforcement learning agent making decisions requires information on not only the propagation state of fake news and true news but also the environmental factors to infer the future state if applying the intensity. In contrast, 
our study aims to select top debunkers in multiple stages in reaction to the dynamic fake news propagation and our proposed reinforcement learning framework minimizes the requirement for information about the environment. 

The Hawkes process~\cite{hawkes1971spectra} and the multivariate Hawkes process~\cite{liniger2009multivariate} have been widely applied in modelling information propagation on social networks \cite{zhou2013learning}  \cite{rizoiu2017hawkes}. They model the propagation either in a self-exciting way \cite{rizoiu2017hawkes} or in a mutual-exciting way \cite{zhou2013learning}. We use the Hawkes process to model the spread of both fake news and true news in a mutual-exciting way. 

\vspace{-0.2cm}
\section{The Multivariate Hawkes Process}\label{sec:hawks}
It is immoral to experiment with real users and spread fake news on real-world social networks, even for research purpose. We use the Multivariate Hawkes process to simulate information propagation on social networks. 
For modeling news propagation on social networks,  the \emph{temporal point process} has been widely used  \cite{zhao2015seismic} \cite{pinto2014modeling}. It can be implemented in neural networks like \emph{recurrent marked temporal point process} \cite{du2016recurrent}, \emph{neural Hawkes process} \cite{mei2016neural} and \emph{neural general temporal point process} \cite{omi2019fully}. In particular, \emph{multivariate Hawkes process} \cite{liniger2009multivariate} is a variant of temporal point process and has been widely applied in fake news research \cite{farajtabar2017fake} \cite{lacombe2018fake} \cite{shu2019studying} \cite{goindani2020cluster} \cite{goindani2020social}.

Briefly, the \emph{Hawkes process} is a stochastic temporal point process model with self-excitement which stimulates the occurrence of a sequence of events. Each event occurrence will excite the process to raise the occurrence probability of the next event \cite{hawkes1971spectra}. Being a counting process, the Hawkes process can be represented as:
\begin{equation}\label{eq1}
N(t)=\sum_{t_{\ell} \leq t} h\left(t-t_{\ell}\right).
\end{equation}
where $t_l$ is the occurring time of $l$-th event, $t$ is the current time, $h(v)$ is the standard Heaviside function such that $h(v) = 1$ if $v >= 0$ and $h(v) = 0$ if $v < 0$ \cite{farajtabar2017fake}. Equation ~\ref{eq1} counts the occurrence of event from time 0 to time $t$. 

To characterise the self-excitement of the Hawkes process, the conditional intensity function is defined to estimate the probability of an event occurrence during an infinitesimal period of time on the condition of history. Formally, 
\begin{equation}\label{eq2}
\lambda(t)=\mu+\sum_{t_{\ell}<t} \phi\left(t-t_{\ell}\right).
\end{equation}
where $\mu$ is the base (background) intensity and $\phi(v)$ is the kernel function. The base intensity is independent of the previous event occurrences while the kernel function is the intensity excited by previous event occurrences. In this paper, we use Hawkes kernel with exponential decay which can be represented as $\phi(v) = \alpha e^{-\omega v}h(v)$ where $\alpha$ is the self-exciting coefficient and $\omega$ is the ratio of kernel decay.

The \emph{Multivariate Hawkes process} (MHP) with $n$ dimensions is applied in the case of $n$ event types. In this setting, the conditional intensity function is defined to estimate the occurrence  probability of event type $i$ during an infinitesimal period of time on the condition of history: 
\begin{equation}\label{eq3}
\lambda_i(t) = \mu_i + \sum_{j=1}^{n} \sum_{t_{j, \ell} < t} \phi_{i, j}\left(t-t_{j, \ell}\right).
\end{equation}
where $t_{j, \ell}$ is the occurring time of $l$-th event of type $j$, $\mu_i$ is the base intensity of type $i$, $\phi_{i,j}(v)$ is the kernel function. MHP expands the Hawkes kernel function from self-excitement to mutual excitement. So, $\phi_{i,j}(v) = \alpha_{ij} e^{-\omega v}h(v)$ where $\alpha_{i,j}\in \mathbf{A}$ is the coefficient indicating to which extent event occurrence of type $j$ influences event occurrence of type $i$. $\mathbf{A}$ is the coefficient matrix. 

The propagation of news -- both fake and true news -- on social networks is modelled by MHP in this study. Specifically, the occurrence of event type $i$ refers to that user $i$ posts a piece of news on social networks. For fake news, the intensity functions are $\boldsymbol{\lambda}^{F}(t) = (\lambda_{1}^{F}(t), \cdots, \lambda_{n}^{F}(t))^{\top}$ with base intensity $\boldsymbol{\mu}^{F}=(\mu_{1}^{F}, \cdots, \mu_{n}^{F})$. For true news, the intensity functions are $\boldsymbol{\lambda}^{M}(t) = (\lambda_{1}^{M}(t), \cdots, \lambda_{n}^{M}(t))^{\top}$ with base intensity $\boldsymbol{\mu}^{M}=(\mu_{1}^{M}, \cdots, \mu_{n}^{M})$. 
 
\section{Problem Statement}\label{sec:problem}
A social network can be modelled as a directed graph $G = (U, E)$ where $U$ denotes graph nodes representing social network users, and $E$ denotes edges between nodes representing the ``following'' relationship between users.~\footnote{We use terms ``network'' and ``graph'', and ``user'' and ``node'' interchangeably.} On graph $G$, the origin of fake news is a set of nodes that spread fake news following the MHP at a given intensity; fake news mitigation is achieved by spreading true news to nodes that received fake news. Without loss of generality, we assume all nodes in $U$ agree to participate in fake news mitigation campaigns.~\footnote{If a subset of nodes $U'\subset U$ agree, the proposed method selects debunkers from $U'$ and no other adaption is required.} Node $i$ in the graph comes with mitigation cost $c^i$. Nodes with more followers will have a higher cost.

A mitigation campaign comprises multiple stages. For the $k$-th stage with budget $l^k$, a number of nodes are selected as {\em debunkers} to spread true news on the network following the MHP, under the constraint that the total mitigation cost of debunkers cannot be over $l^k$.  
The aim of a mitigation campaign is to learn the optimal mitigation policy 
such that the debunkers selected at each stage can maximize the cumulative mitigation effect (or reward) across all stages of the campaign, defined as:
\begin{equation}
\label{eq4}
V^{\pi}\left(s^{0}\right)=\mathbb{E}\left[\sum_{k=0}^{\infty} \gamma^{k} R^{k} \mid s^{0}\right].
\end{equation}
where $s^{0}$ is the propagation state of fake news and true news on the graph at the beginning of the mitigation campaign, $k$ is the identifier of a mitigation stage, and $\gamma^k$ and $R^k$ are the discount factor and the reward at mitigation stage $k$ respectively. 

\section{Methodology}\label{sec:method}
\begin{figure}
    \centering
    \includegraphics[width=0.48\textwidth]{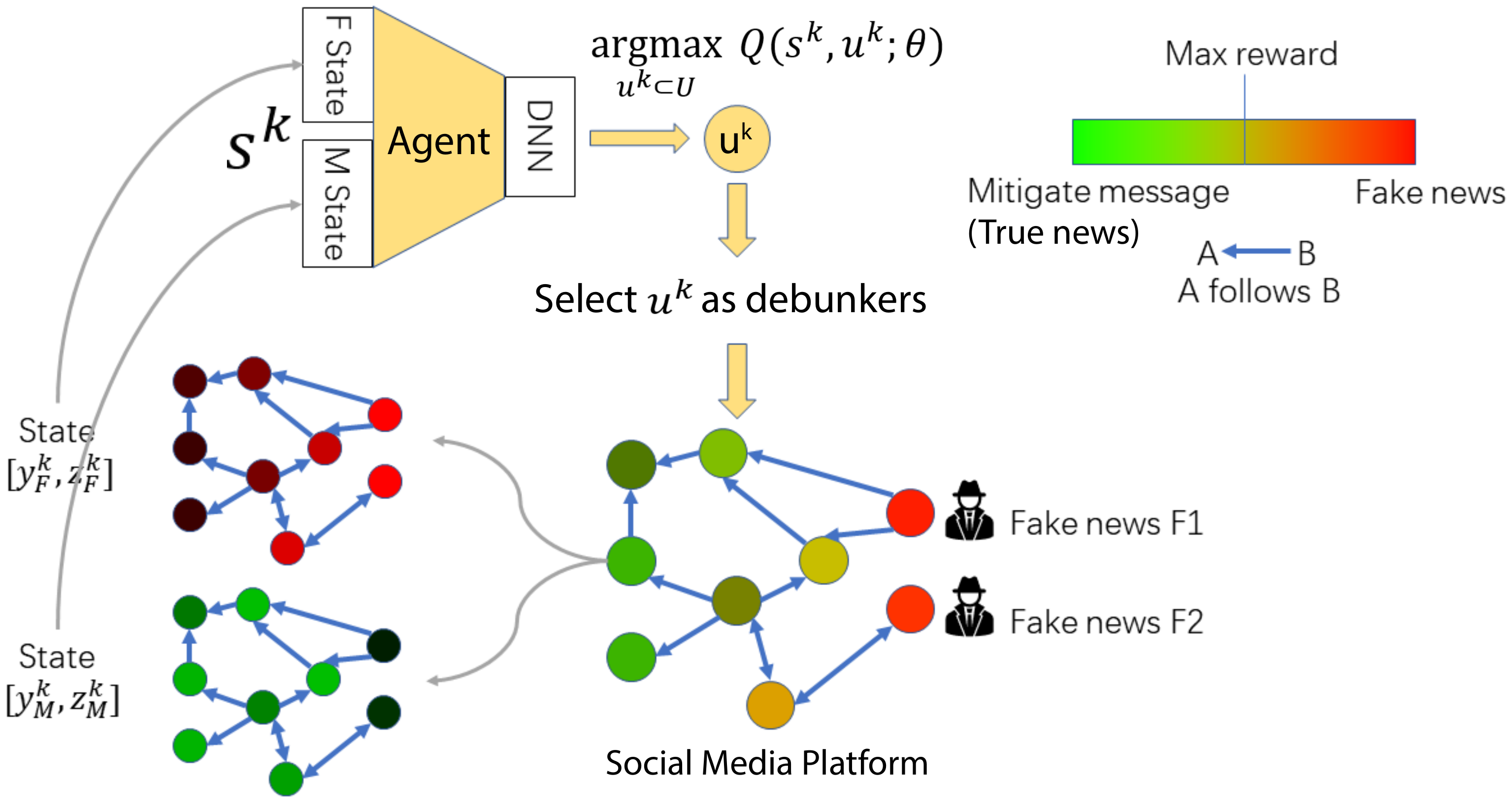}
    \vspace{-0.5cm}
    \caption{The $k$-th stage of a mitigation campaign.}
    \vspace{-0.5cm}
    \label{fig:interact_with_env}
\end{figure}

Given a mitigation campaign of $N$ stages, we apply reinforcement learning to learn the optimal mitigation policy. The proposed framework is called DQN-FSP, namely {\em Deep $Q$-network with Future State Prediction}. 
Figure~\ref{fig:interact_with_env} shows conceptually how DQN-FSP works at stage $k$ of a mitigation campaign. The input of agent is $\textbf{s}^k$ which represents the propagation state of fake news and true news on the (social) network at the beginning of stage $k$. Following the mitigation policy learnt so far, the agent selects a number of nodes $u^k$ as debunkers within budget $l_k$ to perform the $k$-th stage mitigation. At the end of the $k$-th stage mitigation, the reward $R$ is evaluated and the current propagation state of fake news and true news on the network is updated to $s^{k+1}$. The $(k+1)$-th stage mitigation runs in a similar way if $k+1<N$. 

\subsection{State, Action and Reward}\label{sec:rl3}
At the beginning of the $k$-th stage mitigation, the propagation state of fake news (F) and true news (M) on the network, $s^k$, is defined as:
\begin{equation}\label{eq5} 
s^{k}=\left[y_{F}^{k}; y_{M}^{k}; z_{F}^{k}; z_{M}^{k}; e\right].
\end{equation}
where $y_F^k$ and $y_M^k$ are the conditional intensity of users for fake news and true news respectively. Let * be either $M$ or $F$. 
\begin{equation}
y_*^k=(\mathbf{y}_{*, 1}^k, \mathbf{y}_{*, 2}^k, \cdots, \mathbf{y}_{*, n}^k).
\end{equation}
where $\mathbf{y}_{*, i}^k = \sum_{j=1}^{n} \sum_{t_{j, \ell} < t_k} \phi_{i, j}\left(t_k - t_{j, \ell}\right)$ for $1 \leq i \leq n$. It is equivalent to $\lambda_i(t_k)$ without $\mu_i$ (Equation \ref{eq3}), i.e., the conditional intensity of user $i$ at time $t_k$ where $\mu_i$ is ignored since it is a constant. 

In the time period from ($t_k-\Delta_{T}$) to $t_k$, the number of fake news and true news posted by users on the network are represented as $z_F^{k}$ and $z_M^{k}$ respectively. They are components of current propagation state of fake news and true news $s^k$ in Equation \ref{eq5}. Let * be either $M$ or $F$. 
\begin{equation}\label{eq:z}
z_*^k=(\mathbf{z}_{*, 1}^k, \mathbf{z}_{*, 2}^k, \cdots, \mathbf{z}_{*, n}^k).
\end{equation}
where $\mathbf{z}_{*, i}^k = \frac{1}{\Delta_{T}} (N_{i}(t_{k}) - N_{i}(t_{k} - \Delta_{T}))$ for $1 \leq i \leq n$.

In Equation \ref{eq5}, $s^k$ also includes $e$, which is a vector $(\mathbf{e}_1,\cdots,\mathbf{e}_n)$ where $\mathbf{e}_i$ for $1 \leq i \leq n$ is the number of followers of user $i$ in the network. Note that $e$ is the only information about the environment that is required by our method and it is typically available directly from any social network.
It is noteworthy that in previous studies~\cite{farajtabar2017fake}, the reinforcement learning agent requires more information about the environment (such as the ``following'' relationship between users and the coefficient between users) 
to predict future state that the selected action may lead to. In our approach, the future state is predicted using an RNN model (Section~\ref{sec:multiple}) which does not require such environment information. 


\sloppy
Based on input $s^k$, the agent follows the mitigation policy and takes an action such that a set of users $u^k$ are selected as debunkers within stage budget, where arg$\max_{u^k\subset U}Q(s^k,u^k;\theta)$, and $\theta$ is the set of parameters of mitigation policy learnt so far. In other words, given $s^k$, selecting $u^k$ will maximize the expected cumulative reward, or mitigation effect, $Q(s^k,u^k;\theta)$.  


\sloppy
For the $k$-th stage mitigation, the reward is evaluated by \emph{correlation maximization} \cite{farajtabar2017fake}. It is principled on that users exposed to fake news are also exposed to true news. The reward measure is defined as:
\begin{equation}\label{eq6}
r\left(s^{k}, u^{k}\right)=\frac{1}{n} \mathcal{M}^{k}\left(t_{k+1}; s^{k}, u^{k}\right)^{\top} \mathcal{F}^{k}\left(t_{k+1}; s^{k}, u^{k}\right)
\end{equation}
where $\mathcal{M}^{k}\left(*\right) = (\mathcal{M}_{1}^{k}\left(*\right), \cdots, \mathcal{M}_{n}^{k}\left(*\right))$ and $\mathcal{F}^{k}\left(*\right) = (\mathcal{F}_{1}^{k}\left(*\right), \cdots, \mathcal{F}_{n}^{k}\left(*\right))$.
\begin{equation}
\mathcal{M}_{i}^{k}\left(t; s^{k}, u^{k}\right) = \frac{1}{t - t_k}\sum_{j=1}^{n} b_{i j} (N^M_{j}(t) - N^M_{j}(t_{k})).
\end{equation}
\begin{equation}
\mathcal{F}_{i}^{k}\left(t; s^{k}, u^{k}\right) = \frac{1}{t - t_k}\sum_{j=1}^{n} b_{i j} (N^F_{j}(t) - N^F_{j}(t_{k})).
\end{equation}
where $t_k$ is the starting time of the $k$-th mitigation stage, $t$ is the current time, $N^M_{j}(t) - N^M_{j}(t_{k})$ is the number of times that user $j$ spread true news during the time period from $t_k$ to $t$, $N^F_{j}(t) - N^F_{j}(t_{k})$ is the number of times that user $j$ spreads fake news during the time period from $t$ to $t_k$. $\mathbf{B}$ is an adjacency matrix. Given users $i$ and $j$, for $b_{ij}\in \mathbf{B}$, if user $j$ follows user $i$ in the  network we have $b_{ij} = 1$, otherwise $b_{ij} = 0$.    

\subsection{The Multi-stage Mitigation Campaign}\label{sec:multiple}
A mitigation campaign consists of a sequence of mitigation stages. Following some mitigation policy, the agent takes actions to select as many users as possible as debunkers within the budget at each stage so that the cumulative reward for the campaign is maximized. But the search space for debunkers is exponentially large with respect to the total number of users. 
A feasible solution is the greedy strategy of applying some mitigation policy to select debunkers at each stage.
A straightforward policy is to 
select one debunker with the highest reward each time; the policy is then repeatedly applied to select multiple debunkers until the budget for the stage is exhausted. 
But this policy may lead to the issue of mitigation overlap where the true news posted by multiple debunkers are received by the same users. With our model DQN-FSP, we propose the policy to select multiple debunkers.  
We next describe these two policies in detail. 

\subsubsection{A One-debunker Mitigation Policy}\label{sec:single}
At the beginning of the $k$-th stage, a one-debunker mitigation policy modelled as the DQN is applied to select one user $u^k$ ($|u^k|=1$) as the debunker given state $s^k$. That is, the selected user $u^k$ can lead to the maximum $Q(s^k, u^k; \theta)$. At the end of the $k$-th stage, the reward $R^k$ is evaluated and the current propagation state of fake news and true news on the graph is updated to $s^{k+1}$. A new training data instance $(s^k, u^k, R^k, s^{k+1})$ of the one-debunker mitigation policy is created. Specifically, $Q(s^k,u^k;\theta)$ is updated as: 
\begin{equation}\label{eq7}
Q(s^k,u^k;\theta)=\mathbb{E}\left[R^k+\gamma \max _{u^{k+1}} Q\left(s^{k+1}, u^{k+1}; \theta\right) \mid s^k, u^k, \theta\right],
\end{equation}
where $\gamma \in [0, 1]$ is a discount factor to control the influence of future reward. When $\gamma = 1$, all future rewards are fully considered and treated equally. When $\gamma = 0$, only instant reward is considered. The loss used to train the network is calculated as follows:
\begin{equation}\label{eq8}
L(\theta)=\mathbb{E}_{s^k, u^k, R^k, s^{k+1}}\left[\left(y^{DQN}-Q(s^k, u^k; \theta)\right)^{2}\right],
\end{equation}
\begin{equation}
y^{DQN} = \left(R^k + \gamma\max _{u^{k+1}} Q\left(s^{k+1}, u^{k+1}; \theta^{-}\right) \right).
\end{equation}
where $\theta$ is the weight of the online network while $\theta^{-}$ is the weight of the target network, which is updated with the online network regularly. 

The learning procedure of the one-debunker mitigation policy is presented in Algorithm \ref{alg:singelselect} where a single user is selected as the debunker for $k$-th mitigation stage. Note that $U_{legal}^k(l^k)$ is the subset of users whose mitigation cost is less than the stage budget $l^k$. 

\subsubsection{Selecting Multiple Debunkers with DQN-FSP}\label{future_state_predictor}
To minimize the mitigation overlap, we propose an RNN model to predict the future state that the currently selected debunkers may lead to, and thus when the agent selects the next debunker, it will avoid those debunker candidates with overlapping mitigation effect. 

Suppose the one-debunker mitigation policy introduced in Section \ref{sec:single} has been well trained. At the beginning of $k$-th stage, let $H^k$ be the set of debunkers initialized to be empty. We can use the one-debunker mitigation policy to select the first user $u^k_1$ and move it from $U$ to $H^k$. In the same way, we can select the second user $u^k_2$. To avoid the mitigation effect of $u^k_2$ overlapping with that of $u^k_1$, we need to know the propagation state $s^k_1$ of fake news and true news after $u^k_1$ spreads true news. But $s^k_1$ is unknown since the $k$-th mitigation stage does not start yet. 

\begin{figure}
    \centering
    \includegraphics[width=0.3\textwidth]{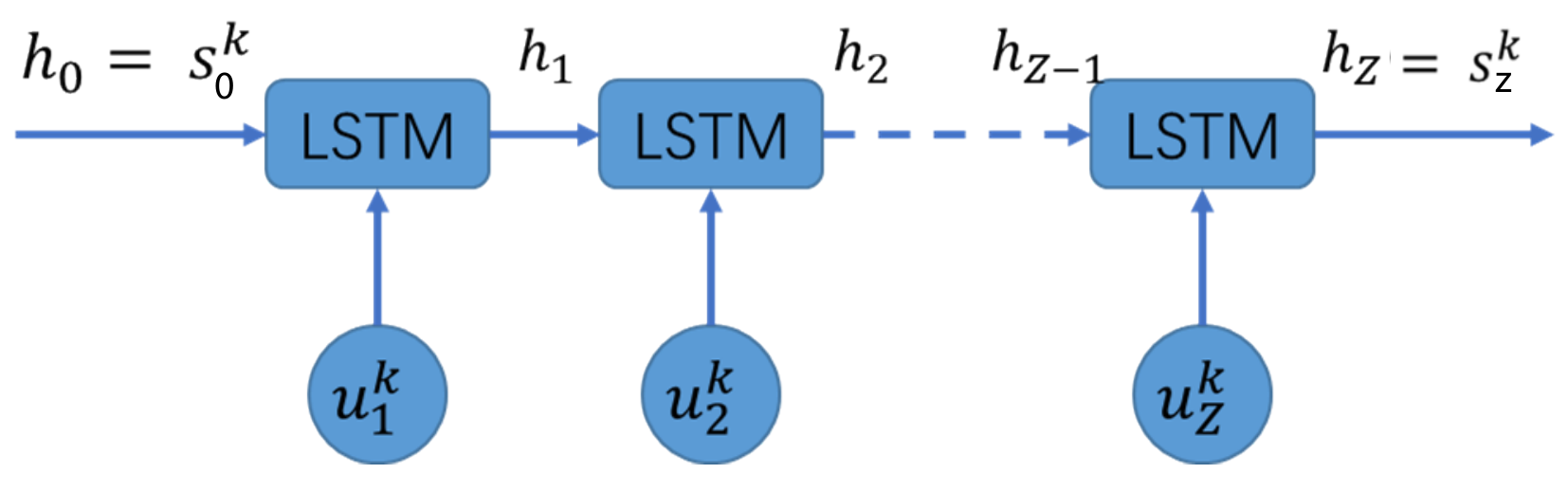}
    \vspace{-0.3cm}
    \caption{Future state prediction with the LSTM RNN model.}
    \label{fig:FSP}
    \vspace{-0.3cm}
\end{figure}

To conquer this problem, we propose a model based on LSTM~\cite{hochreiter1997long} to predict the unknown state $s^k_1$. As shown in Figure \ref{fig:FSP}, LSTM \cite{hochreiter1997long} is used where $s^k_0$ is the initial state of $k$-th mitigation stage and $s^k_z$ is the state after a sequence of debunkers spread true news. For example, we can predict state $s^k_1$ after the first debunker $u^k_1$ spreads true news; with the estimated state $s^k_1$, we can predict $s^k_2$ after the second debunker $u^k_2$ spreads true news, and so on. The pseudo-code of LSTM-based greedy algorithm is presented in Algorithm \ref{alg:multiselect}. 

The training data for the LSTM is collected when training the one-debunker mitigation policy in Section \ref{sec:single}. Given $s^k$ and $u^k$, the training data instance of one-debunker mitigation policy is $(s^k, u^k, R^k, s^{k+1})$. Training data instance of LSTM is $(s^k, u^k, s^{k+1})$. The training process is shown in Algorithm \ref{alg:singelselect}.  

\begin{algorithm}[h]
\small
\caption{Single-debunker Mitigation Policy} 
\label{alg:singelselect}
\begin{algorithmic}[1]
\item Initialize DQN replay memory $D_{DQN}$, FSP memory $D_{fsp}$
\item Initialize action-value function Q with random weights $\theta$
\item Initialize target action-value function $\hat{Q}$ with weights $\theta^{-} = \theta$
\item Initialize FSP with random weights $\theta_{fsp}$
\For{episode = 1, E }
    \State Initialize state $s_0$ and budget for every stage $l^k$;
    \For{k = 1, K}
        \State Observe environment to obtain state $s_k$;
        \State Select action $u^k = argmax_{u \in U_{legal}^k(l^k)} Q(s^k,u; \theta)$;
        \State Perform the single-debunker $u^k$ mitigation task; 
        \State Observe reward $r^k$ and updated state $s^{k+1}$;
        \State Store $(s^k, u^k, r^k, s^{k+1})$ in $D_{DQN}$;
        \State Store $(s^k, u^k, s^{k+1})$ in $D_{fsp}$;
        \State Update $\theta$ using sampled minibatch from $D_{DQN}$;
        \State Every $C$ steps reset $\hat{Q} = Q$;
\EndFor
    \State Update $\theta_{fsp}$ using sampled minibatch from $D_{fsp}$;
\EndFor 
\end{algorithmic} 
\end{algorithm}
\vspace{-0.3cm}

\vspace{-0.3cm}
\begin{algorithm}[h]
\small
\caption{Multi-debunker Mitigation with DQN-FSP} 
\label{alg:multiselect}
\begin{algorithmic}[1] 
\item Initialize FSP memory $D_{fsp}$
\item Initialize action-value function Q with weights $\theta$ trained from single node selection
\For{episode = 1, E }
    \State Initialize state $s_0$, budget for every stage $l^k$
    \For{k = 1, K}
        \State Initialize $H^k$ empty;
        \State Observe state $s_k$ and set hidden state LSTM to be $s^k$ ;
        \While {$ \exists u_i \in U, c^i < l^k $} 
            \State Select action $u^k = argmax_{u \in U_{legal}^k(l^k)} Q(s^k,u; \theta)$;
            \State Feed $u^k$ into LSTM, obtain predicted $s^{k'}$;
            \State Update $s^k = s^{k'}$, $l^k = l^k - c^{u_k}$;
        \EndWhile 
        \State Perform multi-debunkers $H^k$ mitigation task;
        \State Observe reward $r^k$ and updated state $s^{k+1}$;
        \State Store transition $(s^k, H^k, s^{k+1})$ in $D_{fsp}$;
\EndFor
    \State Update $\theta_{fsp}$ using sampled minibatch from $D_{fsp}$;
\EndFor 
\end{algorithmic} 
\end{algorithm}
\vspace{-0.3cm}


\section{Experiments}\label{experiments}
We evaluated our DQN-FSP model on synthetic data with controlled settings and real-world social network data. We ran our experiments on a Slurm cluster consisting of 4 CPU nodes (2 x Intel Xeon E5-2450L, 64G Ram) and 1 GPU node (2 x Intel Xeon E5-2650 v2, 64G Ram, 2 x NVIDIA Tesla M40). All deep networks including DQN and LSTM are implemented in the Tensorflow framework and MHP is implemented using the Tick package~\cite{bacry2017tick}.

\subsection{Baselines}\label{subsec:baseline}
DQN-FSP is compared against six baselines, including:
\begin{itemize}
	\item \emph{Random} (RND): This method is used as a sanity check for all other models, including the baselines. At each stage, it randomly selects a set of nodes as debunkers within budget. 
	\item \emph{Max Influence} (MAX-INF): This method is based on the policy of selecting nodes with the maximal influence~\cite{saxena2020mitigating} \cite{saxena2020k}. For node $i$, the influence is $p_i^k = \mathbf{z}_{M, i}^k \mathbf{z}_{F, i}^k$ where $\mathbf{z}_{*, i}^k$ is as defined in Section \ref{sec:rl3}. 
	\item \emph{Max Coverage} (MAX-COV): This method is an intuitive baseline that maximizes the number of mitigation nodes within budget at each stage. So this method sorts nodes by their mitigation cost and selects the cheapest node first.
	\item \emph{Neural Network} (NN): This method is a learning-based baseline. A classifier is trained. For each training data instance, the current propagation state of fake news and true news is input, and output is the direct mitigation reward (Equation \ref{eq6}) after selecting this node as the debunker to spread true news. At each stage, the trained classifier estimates the direct reward for each node and selects a set of nodes within budget with the highest direct rewards. 
	\item \emph{Deep Q-Network} (DQN): As described in Section \ref{future_state_predictor}, this is the straightforward implementation of the one-debunker mitigation policy to select multiple debunkers at each stage. Different from baseline NN, the objective of selection is to maximize cumulative reward rather than the direct reward. 
	\item \emph{Least-squares Temporal Difference} (LTD): This method follows the idea of existing studies where the same set of debunkers are applied at different stages throughout a mitigation campaign \cite{farajtabar2017fake}.
\end{itemize}
	
	

\begin{figure*}
    \centering
\subfigure[Density test] {
 \label{fig:syna}     
\includegraphics[width=0.45\columnwidth]{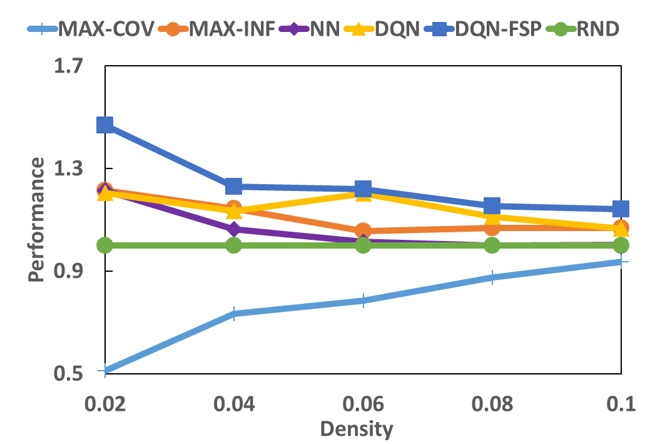} 
}     
\subfigure[Network size test] { 
\label{fig:synb}     
\includegraphics[width=0.45\columnwidth]{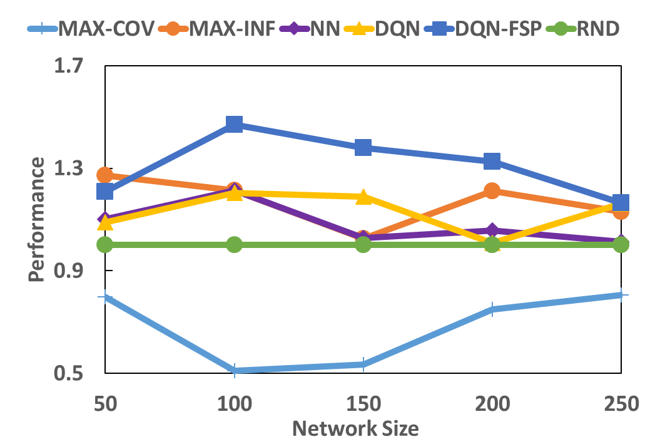}     
} 
\subfigure[Average stage length test] { 
\label{fig:sync}     
\includegraphics[width=0.45\columnwidth]{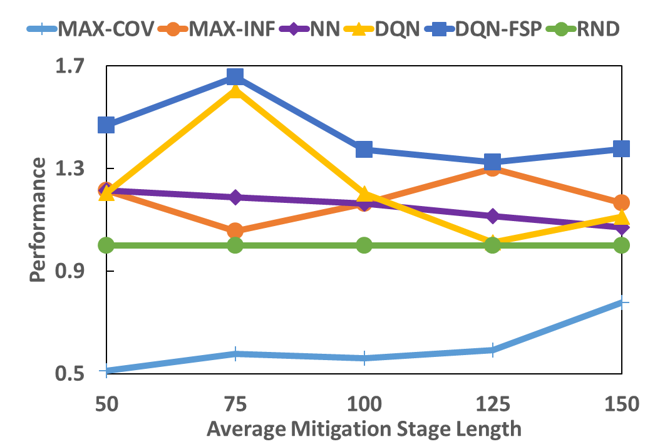}     
} 
\subfigure[Number of stages test] { 
\label{fig:synd}     
\includegraphics[width=0.45\columnwidth]{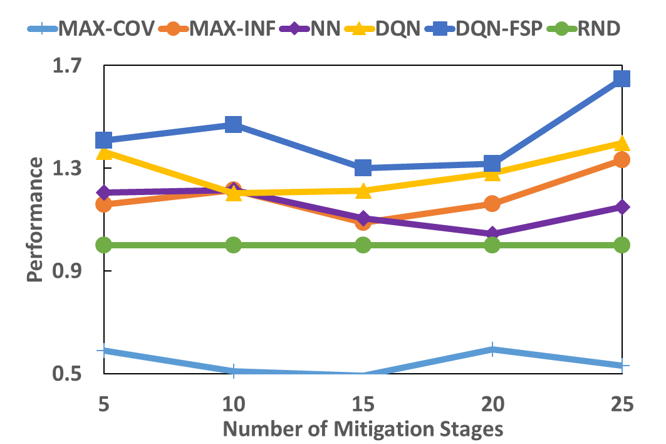}     
}   
\vspace{-0.5cm}
    \caption{Performance with respect to different settings on synthetic test.}
\vspace{-0.3cm}
    \label{fig:syn-param}
\end{figure*}

\subsection{Performance on synthetic data}\label{sec:synthetic}

\subsubsection{Parameter Settings}\label{sec:ses} 
Unless stated otherwise, the graph has $n = 100$ nodes. Between any two nodes, the edge was generated with probability $0.02$. The parameters in Equation \ref{eq3} were set as follows. The coefficient matrix $\mathbf{A}$ was set as $\mathbf{A_{fc}} \odot \mathbf{B}, \mathbf{A_{fc}} = a_{ij} \sim \mathcal{U}[0,0.5]$ where $\mathbf{A_{fc}}$ is the coefficient matrix for all nodes and $\mathbf{B}$ is the adjacency matrix (see Section \ref{sec:rl3}). The coefficient matrix $\mathbf{A}$ was scaled such that the spectral radius is $0.8$ to keep MHP stable. The parameter $\omega$ of kernel function was set to $1$. The base intensity of mitigation was set to $\mu^{M} \sim \mathcal{U}[0,0.1]$ and the base intensity of fake news was set to $\mu^{F} \sim \mathcal{U}[0,0.2]$ to simulate scenarios that fake news already exist on social networks.  

In the default environment settings, we assume that 5 nodes are the fake news spreaders and all the other nodes can be selected as debunkers. A mitigation campaign spans a time window of size 500 and has $10$ mitigation stages. The starting time of each mitigation stage is a random value in [0, 500]. For each node, the mitigation cost is a value in $[1, 5]$. A node has a higher mitigation cost if having more followers. For each mitigation stage, the budget $l^k$ is a random value in $[5, 50]$. Once a node is selected as a debunker, the intensity of the node to spread true news is increased by 3. 

For the cumulative mitigation reward in Equation \ref{eq7}, the discount factor $\gamma = 0.8$. For Equation \ref{eq:z}, $\Delta_{T} = 25$. In experiments, 200 mitigation campaigns (i.e., 200 episodes) are directly processed by the baselines \emph{RND}, \emph{MAX-INF} and \emph{MAX-COV}; for \emph{NN}, \emph{DQN} and \emph{DQN-FSP}, 100 of them are training data and other 100 are test data.

\begin{figure*}
    \centering
\subfigure[Density test on GUR] {
 \label{fig:reala}     
\includegraphics[width=0.45\columnwidth]{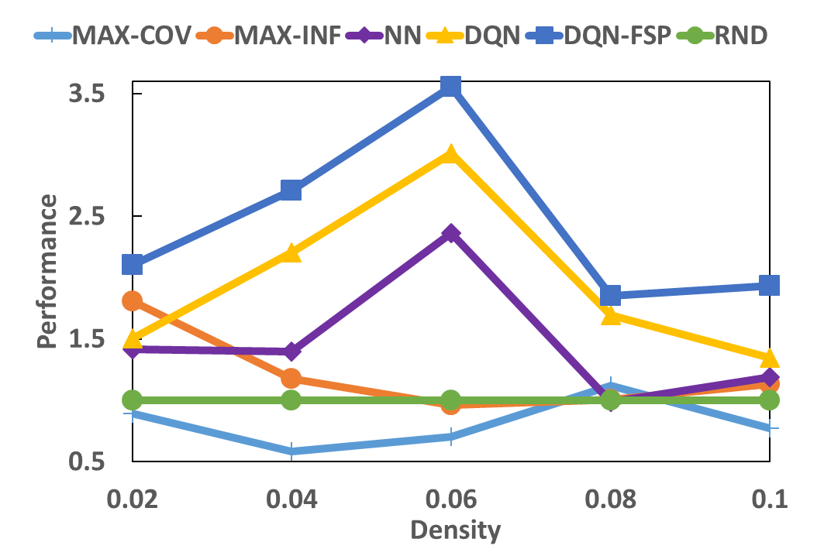} 
}     
\subfigure[Average stage length test on GUR] { 
\label{fig:realb}     
\includegraphics[width=0.45\columnwidth]{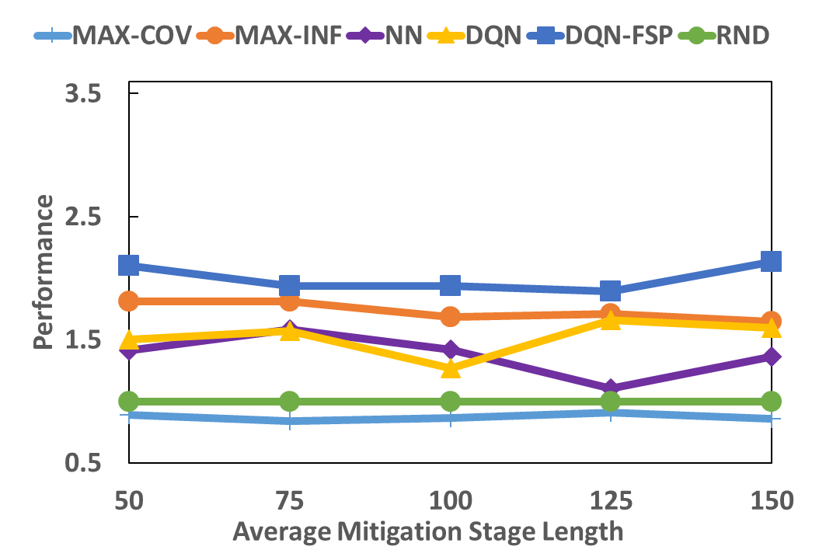}     
}    
\subfigure[Number of stages test on GUR] { 
\label{fig:realc}     
\includegraphics[width=0.45\columnwidth]{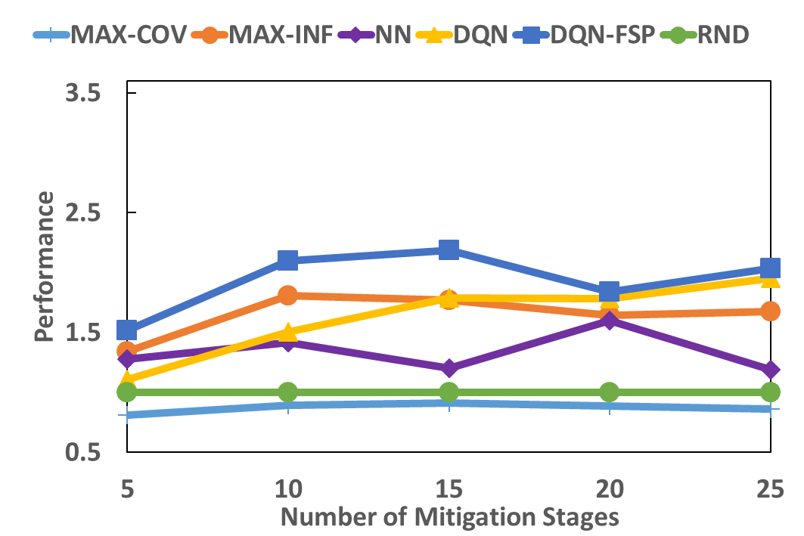}     
} 
\\
\subfigure[Density test on PRI] {
 \label{fig:reald}     
\includegraphics[width=0.45\columnwidth]{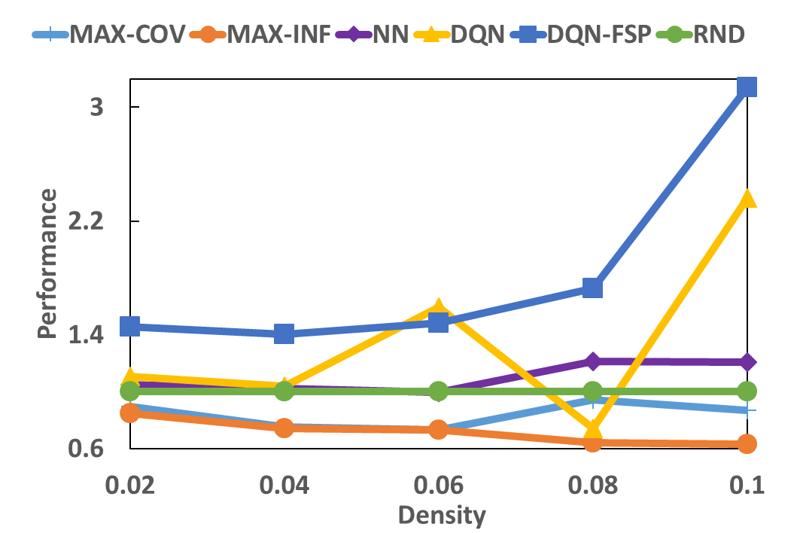} 
}     
\subfigure[Average stage length test on PRI] { 
\label{fig:reale}     
\includegraphics[width=0.45\columnwidth]{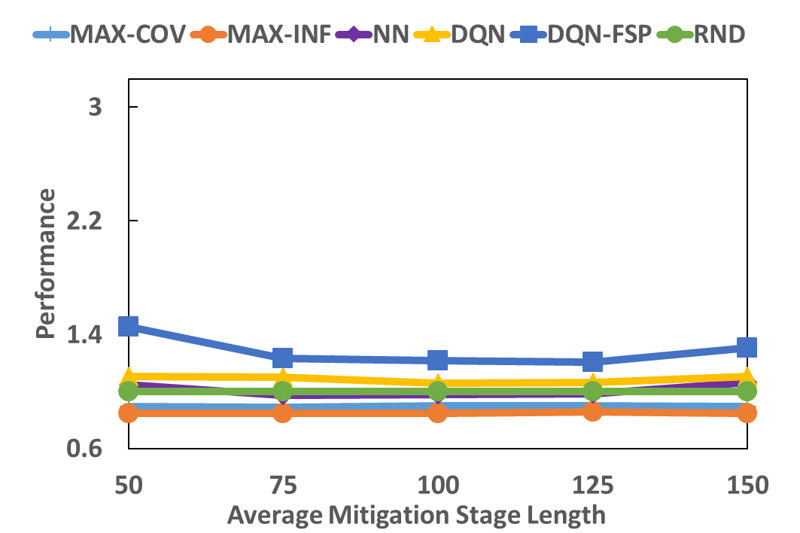}     
}    
\subfigure[Number of stages test on PRI] { 
\label{fig:realf}     
\includegraphics[width=0.45\columnwidth]{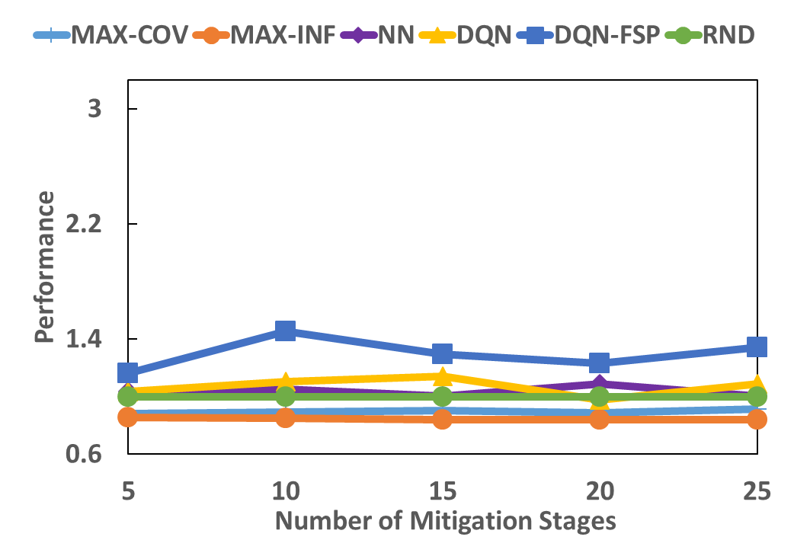}     
} 
\\
\subfigure[Density test on PUT] {
 \label{fig:realg}     
\includegraphics[width=0.45\columnwidth]{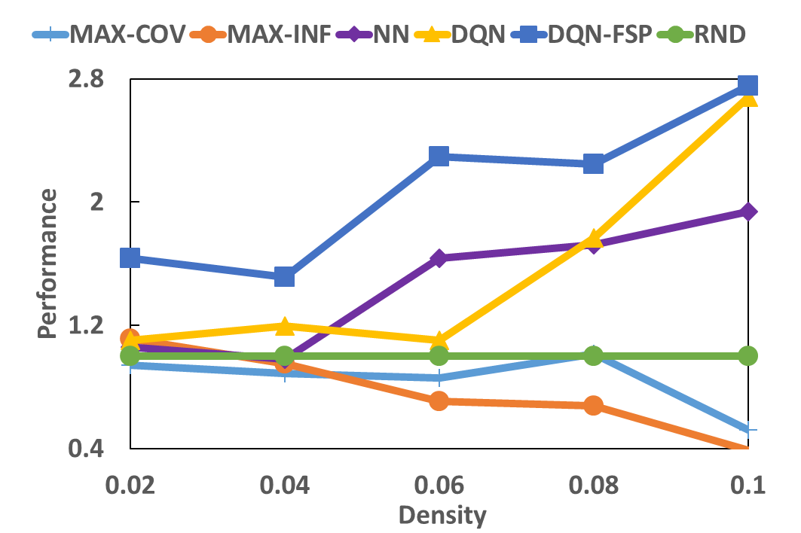} 
}     
\subfigure[Average stage length test on PUT] { 
\label{fig:realh}     
\includegraphics[width=0.45\columnwidth]{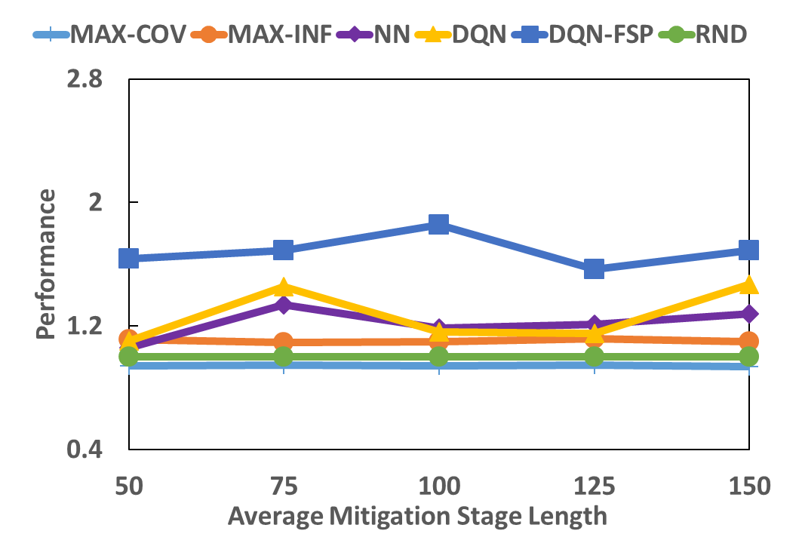}     
}    
\subfigure[Number of stages test on PUT] { 
\label{fig:reali}     
\includegraphics[width=0.45\columnwidth]{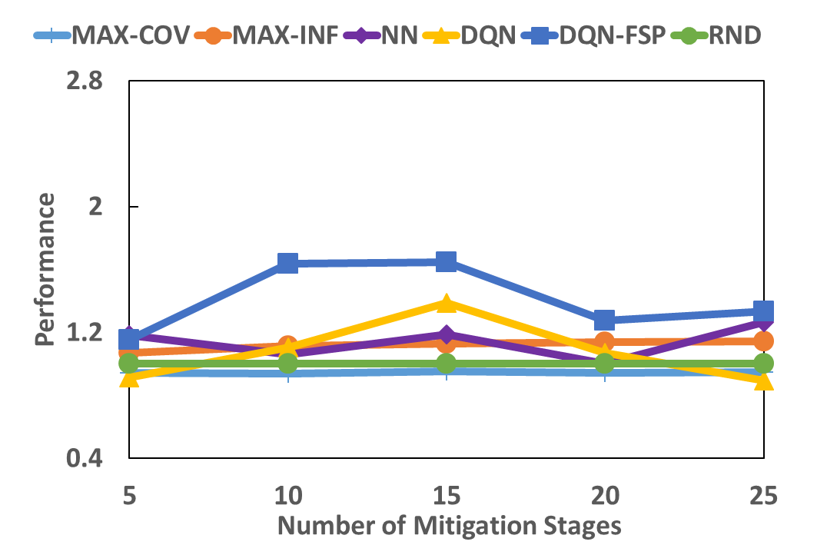}     
} 
\vspace{-0.5cm}
    \caption{Performance with respect to different settings on real-world data.}
\vspace{-0.3cm}
    \label{fig:real-param}
\end{figure*}

\subsubsection{Impact of Environment settings}
The performance is measured by the average cumulative mitigation effect using our DQN-FSP and baselines on test data (3 runs on 100 mitigation campaigns) where parameters are randomly set in the given ranges as discussed in Section \ref{sec:ses}. In Figure \ref{fig:syn-param}, using \emph{RND} as the benchmark, the performance of our DQN-FSP and other baselines are presented as the ratio against the benchmark. 

Figure \ref{fig:syn-param}(a) shows the performance with respect to the density of the social network. By default, a social network has 100 nodes and the edge was generated with a probability of $0.02$ between any two nodes. The probability of edge generation increases from $0.02$ to $0.1$ to simulate different levels of density. We can observe the significant advantage of our DQN-FSP against all baselines at different density levels. In particular, DQN-FSP outperforming \emph{DQN} shows the effectiveness of our future state prediction strategy. While others have declining performance when the social network becomes more dense (i.e., density changes from 0.02 to 0.1), the performance of \emph{MAX-COV} keeps increasing. This confirms previous findings~\cite{farajtabar2017fake} that if the social network is highly dense, no matter which nodes are selected as debunkers, all nodes will be exposed to true news. But in the real world, social networks are usually sparse. In addition, the experiment results show that \emph{MAX-COV} has clearly worse performance compared with \emph{RND} at different settings, that is, selecting cheapest nodes (i.e., with least followers) is worse than selecting nodes randomly. 

Figure \ref{fig:syn-param}(b) shows the performance with respect to the social network size. By default, the social network has 100 nodes and the edge was generated with a probability of $0.02$ between any two nodes. The number of nodes changes from 50 to 250 to simulate different sizes of the social network. Our DQN-FSP model demonstrated consistently better performance than all baselines. 


\subsubsection{Impact of Mitigation Settings}
Figure \ref{fig:syn-param}(c) shows the performance with respect to the average stage length (size of the time window of the mitigation campaign changes accordingly). Longer stage length means the actions taken by the agent at the beginning of the stage will last longer and thus have more effect on the environment.
From the figure, we can see that the proposed method performs well in all settings.

Figure \ref{fig:syn-param}(d) shows the performance of different models with respect to the number of stages in the mitigation campaign. More stages will have more opportunities to optimize mitigation according to the propagation state of fake news and true news at the time and will have more chances to increase the intensity for selected debunkers to spread true news. The results have validated that the performance tends to be improved with an increasing number of stages. With an increasing number of stages, the performance of most methods against \emph{RND} has been improved. At different settings, DQN-FSP outperforms all other baselines.




\subsection{Performance on real-world data}\label{sec:realworld}
\subsubsection{Dataset and Settings}
PHEME~\cite{zubiaga2016analysing}, a widely used dataset for rumour spread on Twitter, has been used in our experiments. It includes the source and timestamp of  Twitter messages -- who post the tweets and when -- and the spread of messages -- who retweets -- for three news topics, including ``Gurlitt" (GUR), ``Prince Toronto" (PRI) and ``Putin Missing" (PUT). Statistics about the dataset is shown in Table~\ref{tbl:realdata}. 

We have used the data to learn the environment parameters $(\mathbf{A}, \boldsymbol{\mu})$ using least square loss. The coefficient matrix was scaled so that the spectral radius is 0.8 to keep MHP stable. Since the dataset does not have data for the social network, we generate edges for the network in different settings to test the robustness of the proposed method. Unless stated otherwise, the network density is set to 0.02. The decay of kernel function is set as $\omega = 1$. Once a user is selected as a debunker, the mitigation intensity is increased by $1$. Mitigation stage and other settings follow the settings for the synthetic data.

\begin{table}
    \small
	\setlength{\abovecaptionskip}{0pt}
	\setlength{\belowcaptionskip}{0pt}
	\caption{Statistics of the real-world PHEME dataset}
	\centering
	\begin{tabular}{c||c|c|c}
		\hline
		\textbf{Topic} & \textbf{\#Users} & \textbf{Fake tweets} & \textbf{
		True tweets}  \\
		\hline
		Gurlitt (GUR) & 98 & 70  & 159 \\  \hline
		Prince Toronto (PRI) & 322 & 483 & 489 \\ \hline
		Putin Missing (PUT) & 352 & 251  & 468 \\ \hline
	\end{tabular}\label{tbl:realdata}
	\vspace{-0.5cm}
\end{table}

\begin{figure*}
    \centering
\subfigure[Background] {
 \label{fig:act1bg}     
\includegraphics[width=0.32\columnwidth]{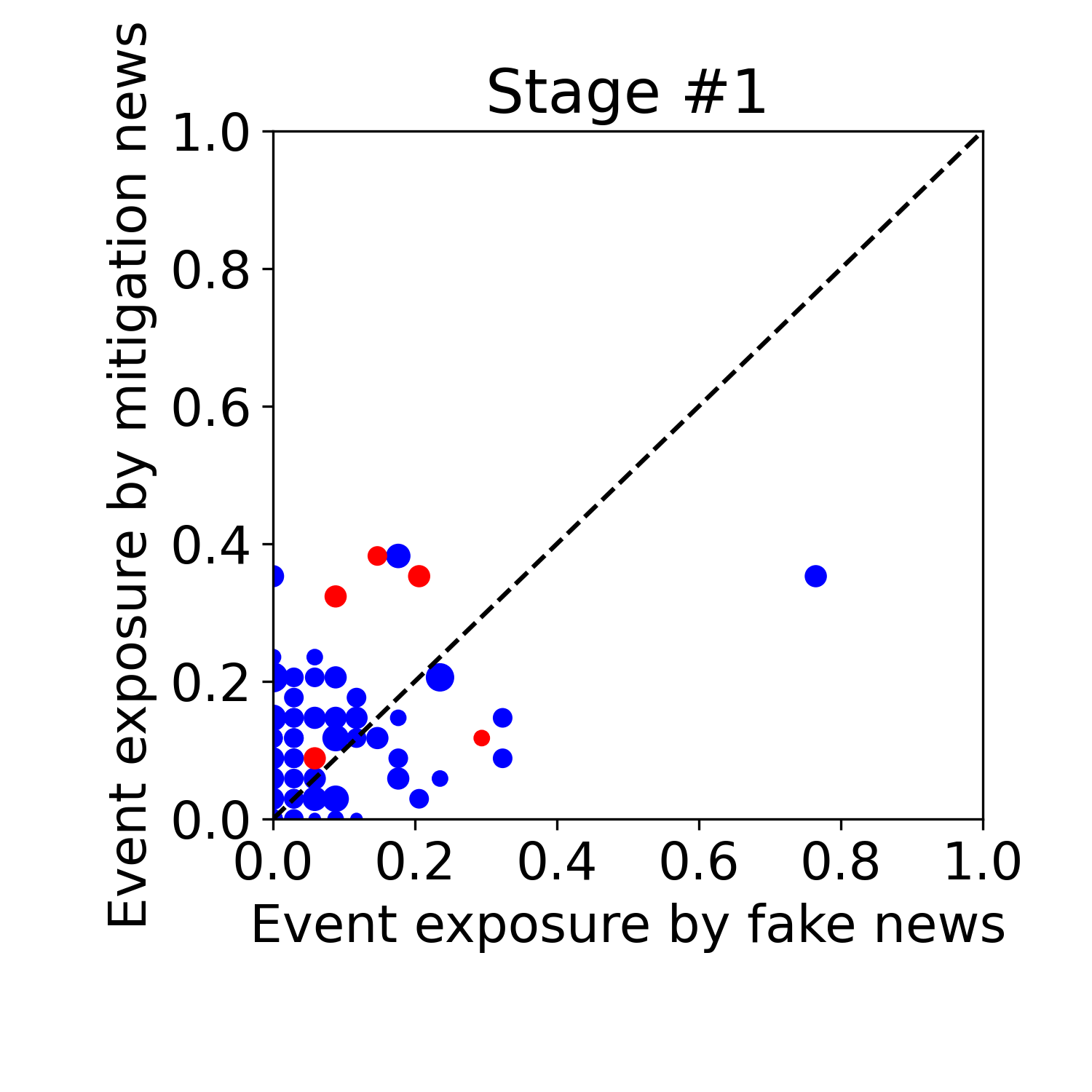} 
} 
\subfigure[Background] { 
\label{fig:act5bg}     
\includegraphics[width=0.32\columnwidth]{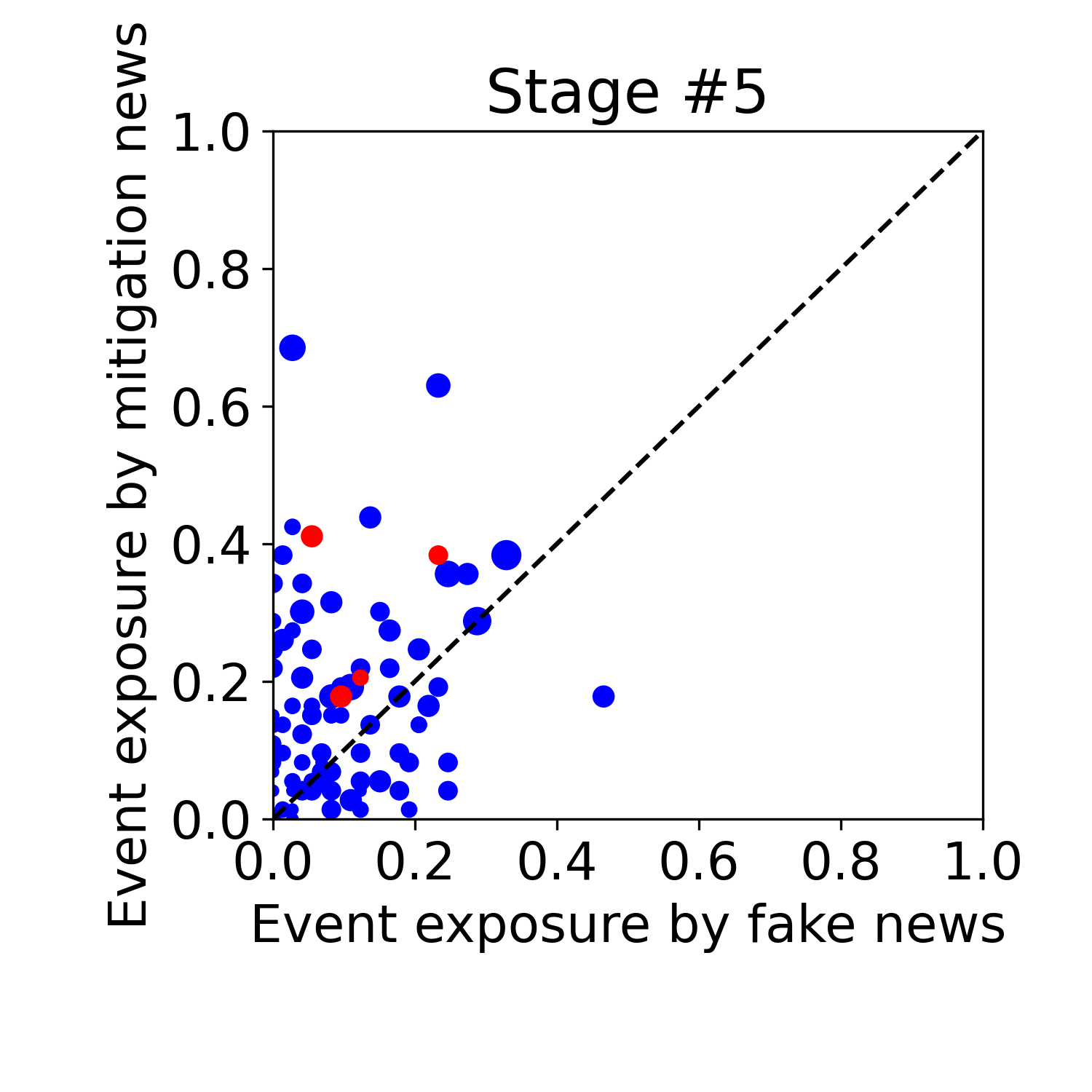}     
}
\subfigure[Background] { 
\label{fig:act9bg}     
\includegraphics[width=0.32\columnwidth]{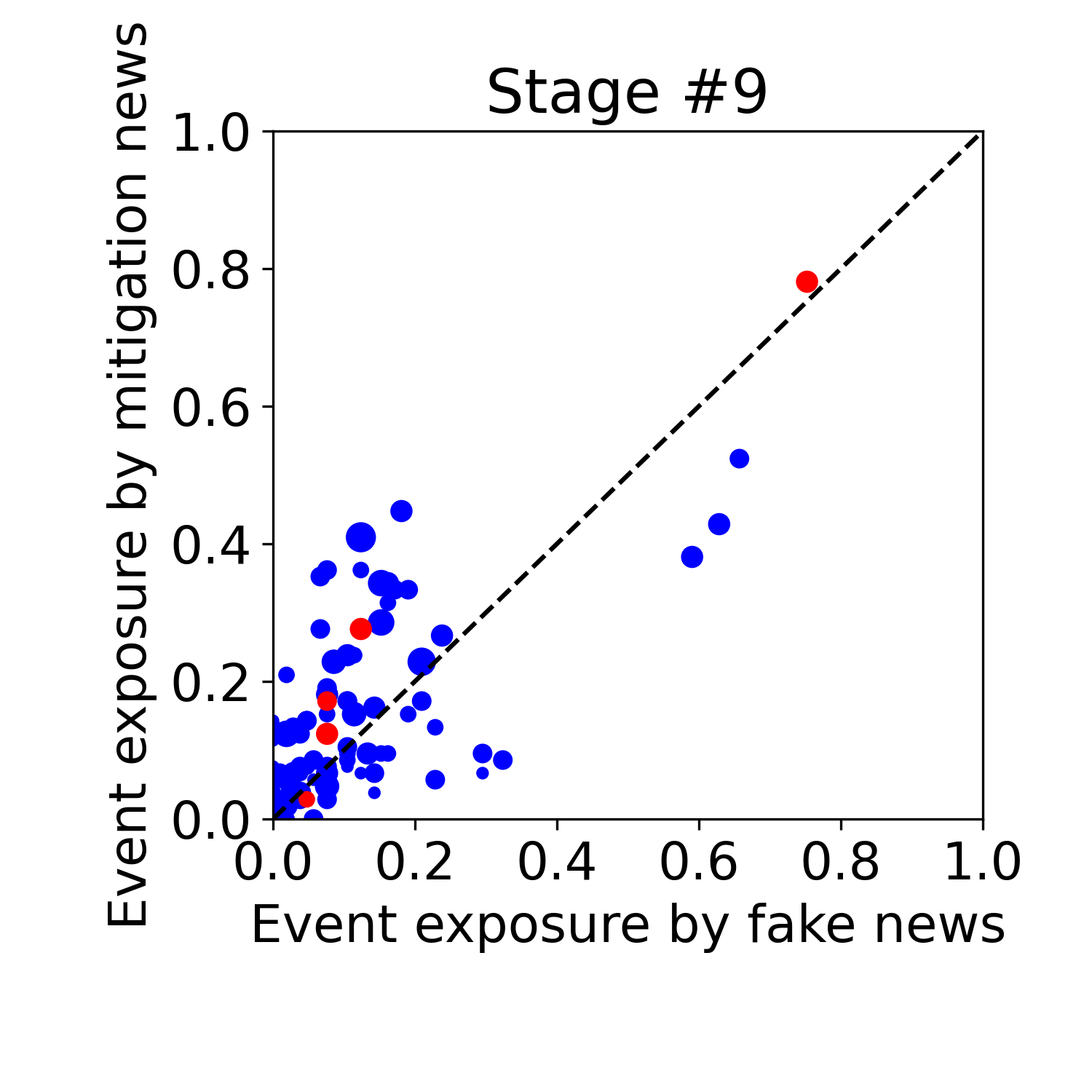}     
} 
\subfigure[Budget 5] {
 \label{fig:act1b5}     
\includegraphics[width=0.32\columnwidth]{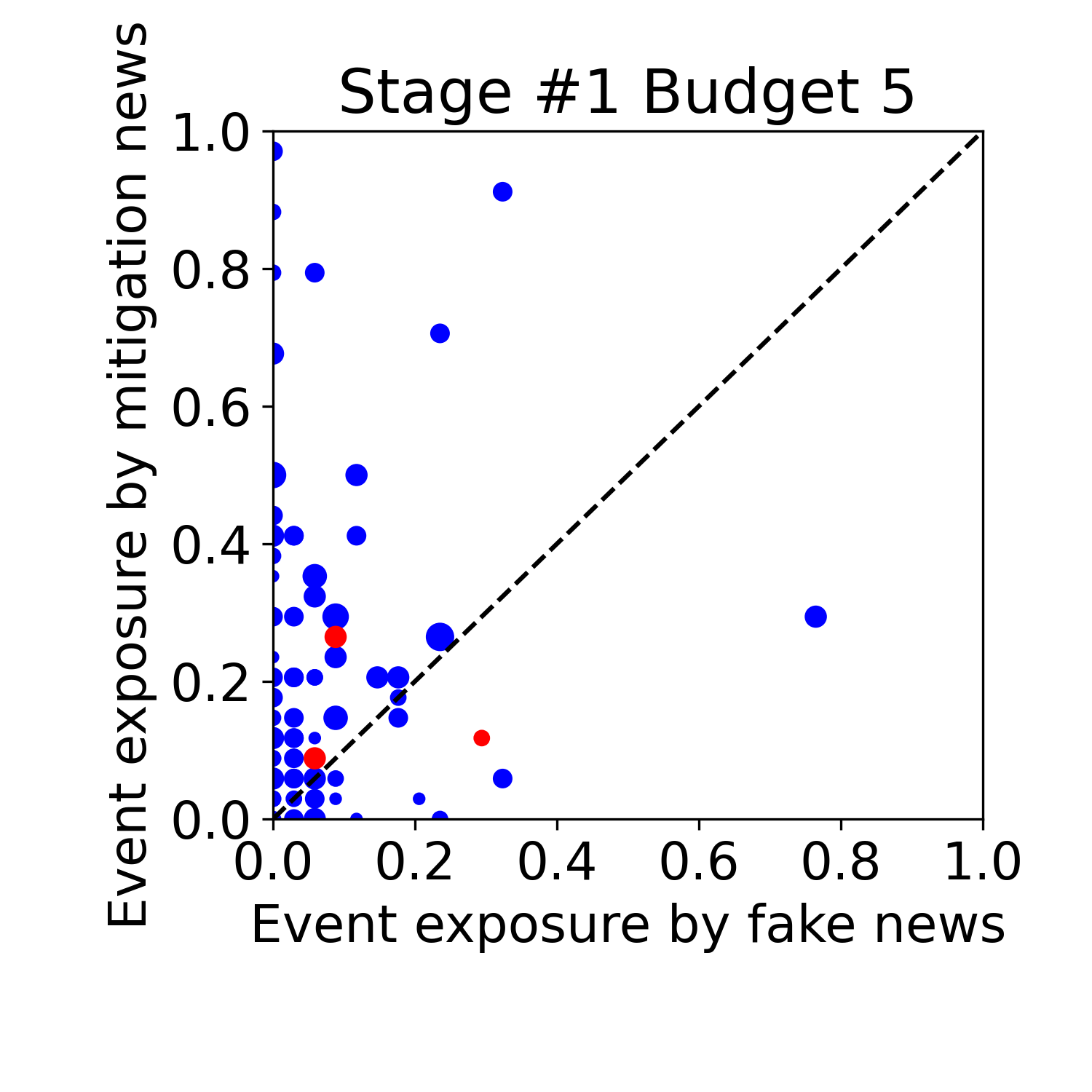} 
} 
\subfigure[Budget 5] { 
\label{fig:act5b5}     
\includegraphics[width=0.32\columnwidth]{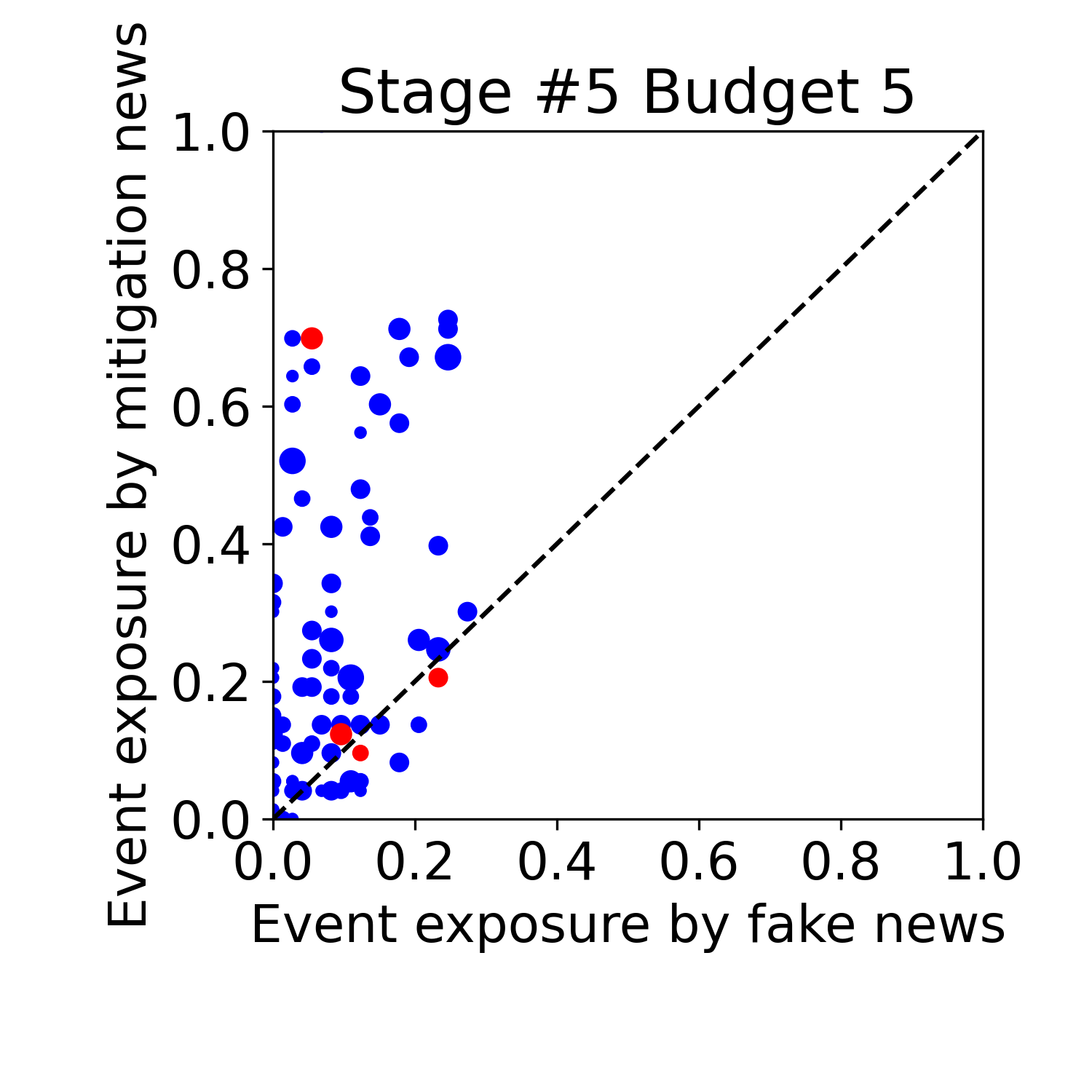}     
}    
\subfigure[Budget 5] { 
\label{fig:act9b5}     
\includegraphics[width=0.32\columnwidth]{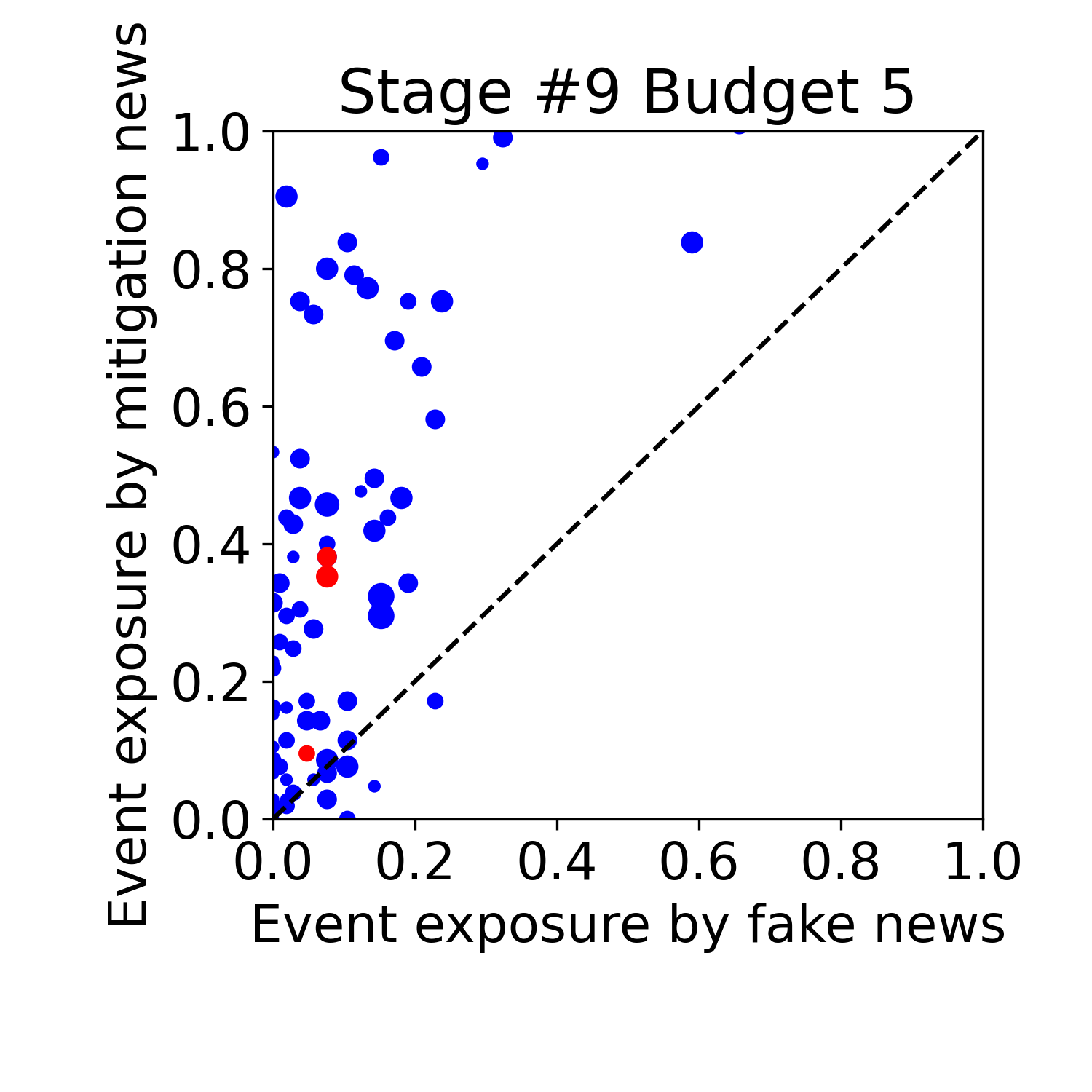}     
}

\subfigure[Budget 10] {
 \label{fig:act1b10}     
\includegraphics[width=0.32\columnwidth]{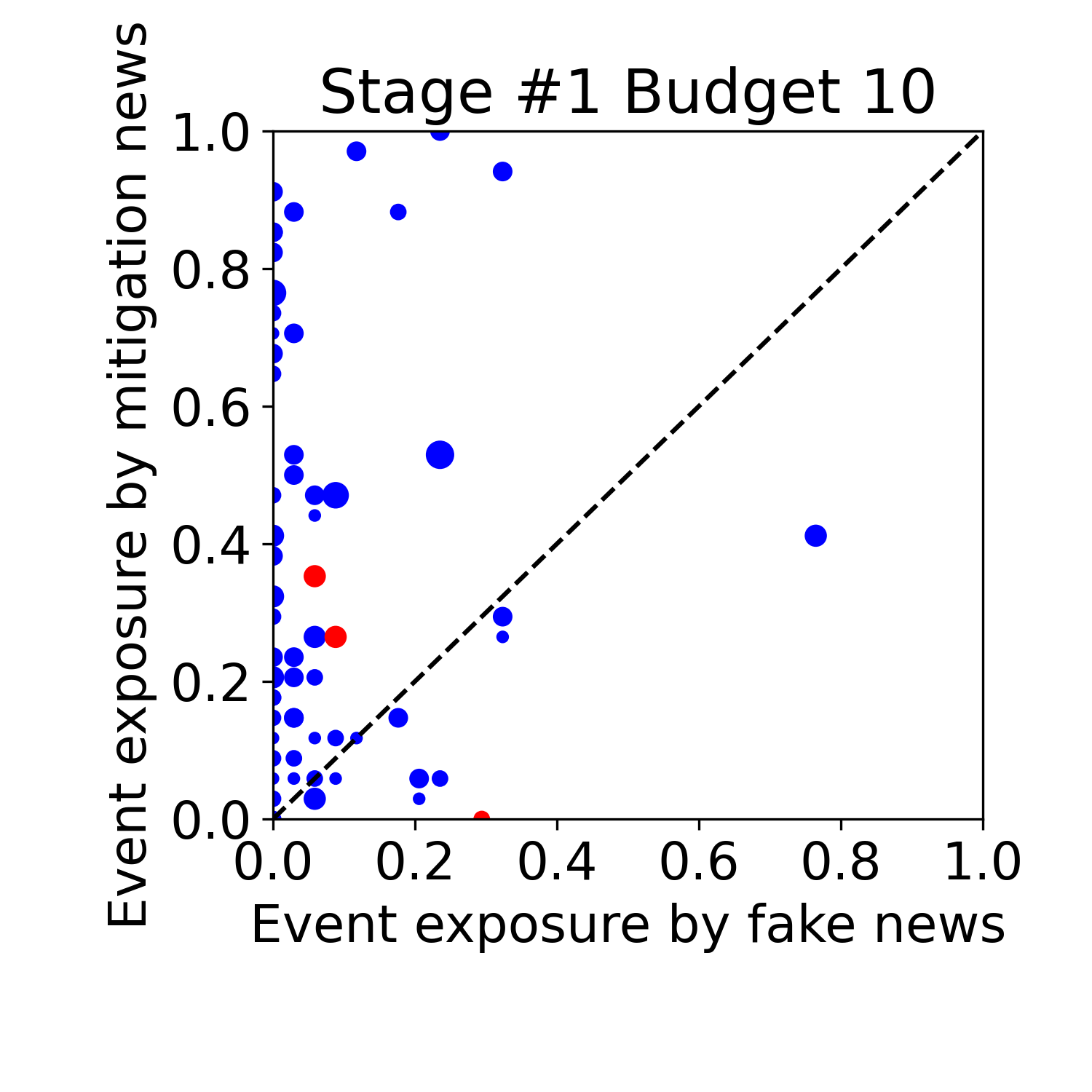} 
} 
\subfigure[Budget 10] { 
\label{fig:act5b10}     
\includegraphics[width=0.32\columnwidth]{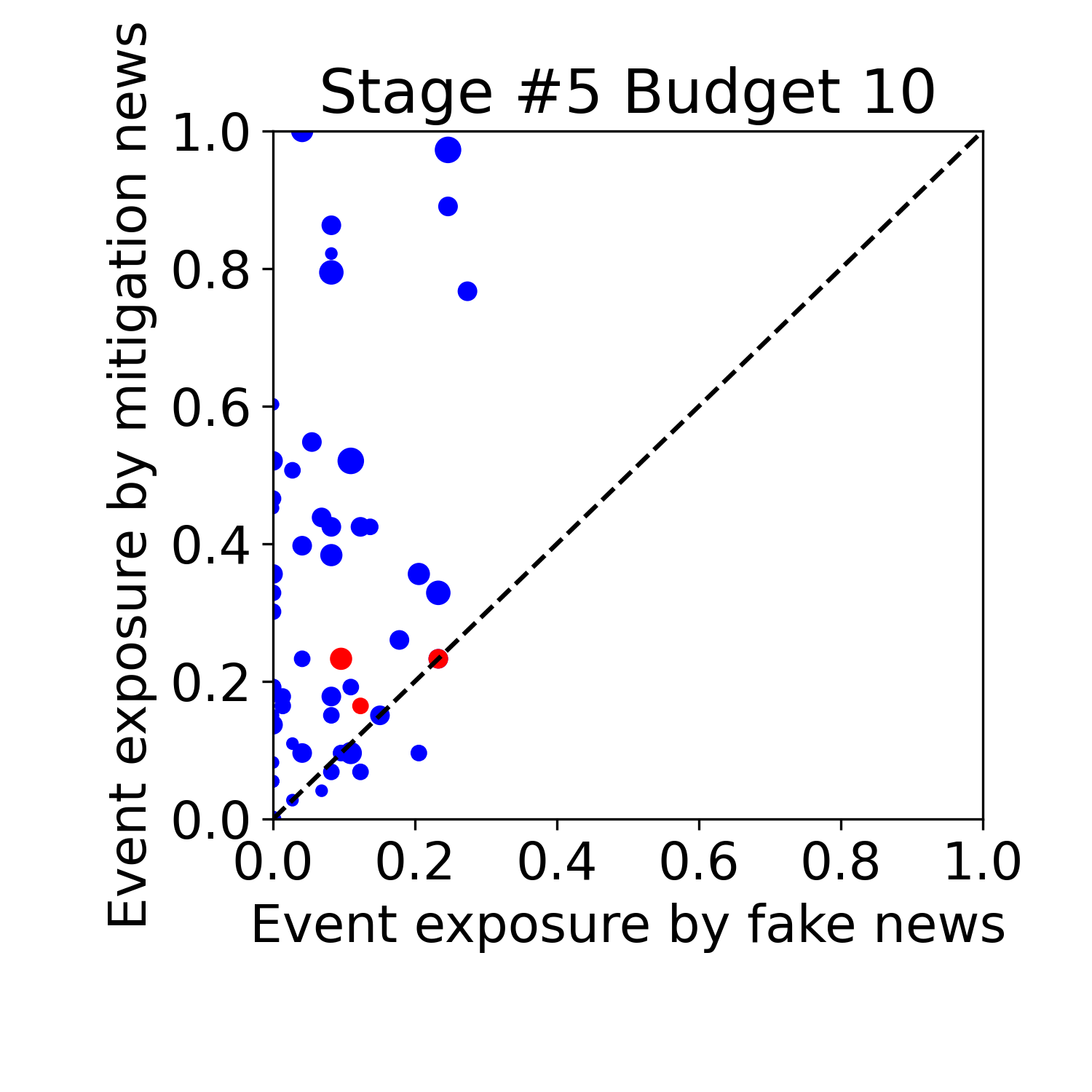}     
}
\subfigure[Budget 10] { 
\label{fig:act9b10}     
\includegraphics[width=0.32\columnwidth]{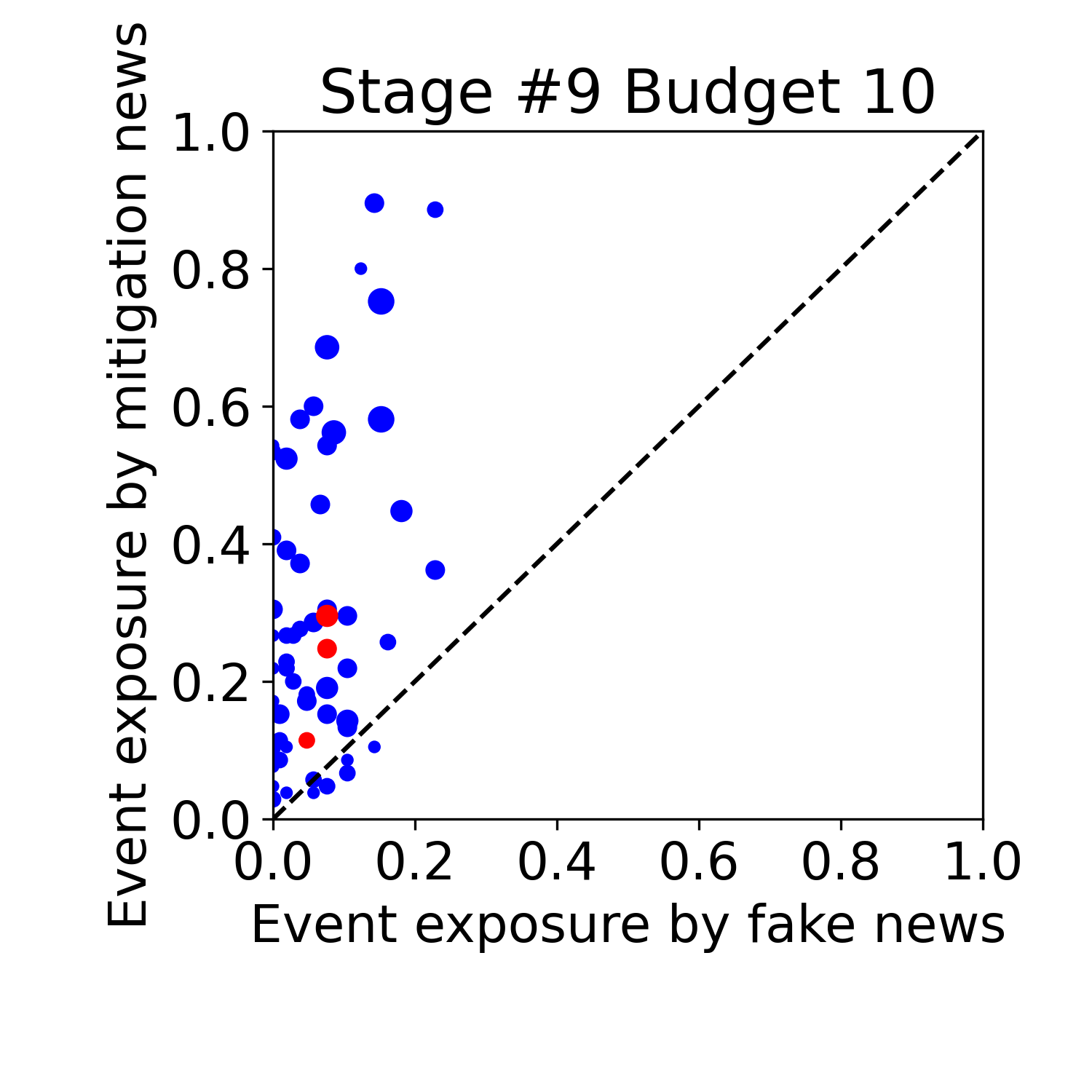}     
} 
\subfigure[Budget 20] {
 \label{fig:act1b20}     
\includegraphics[width=0.32\columnwidth]{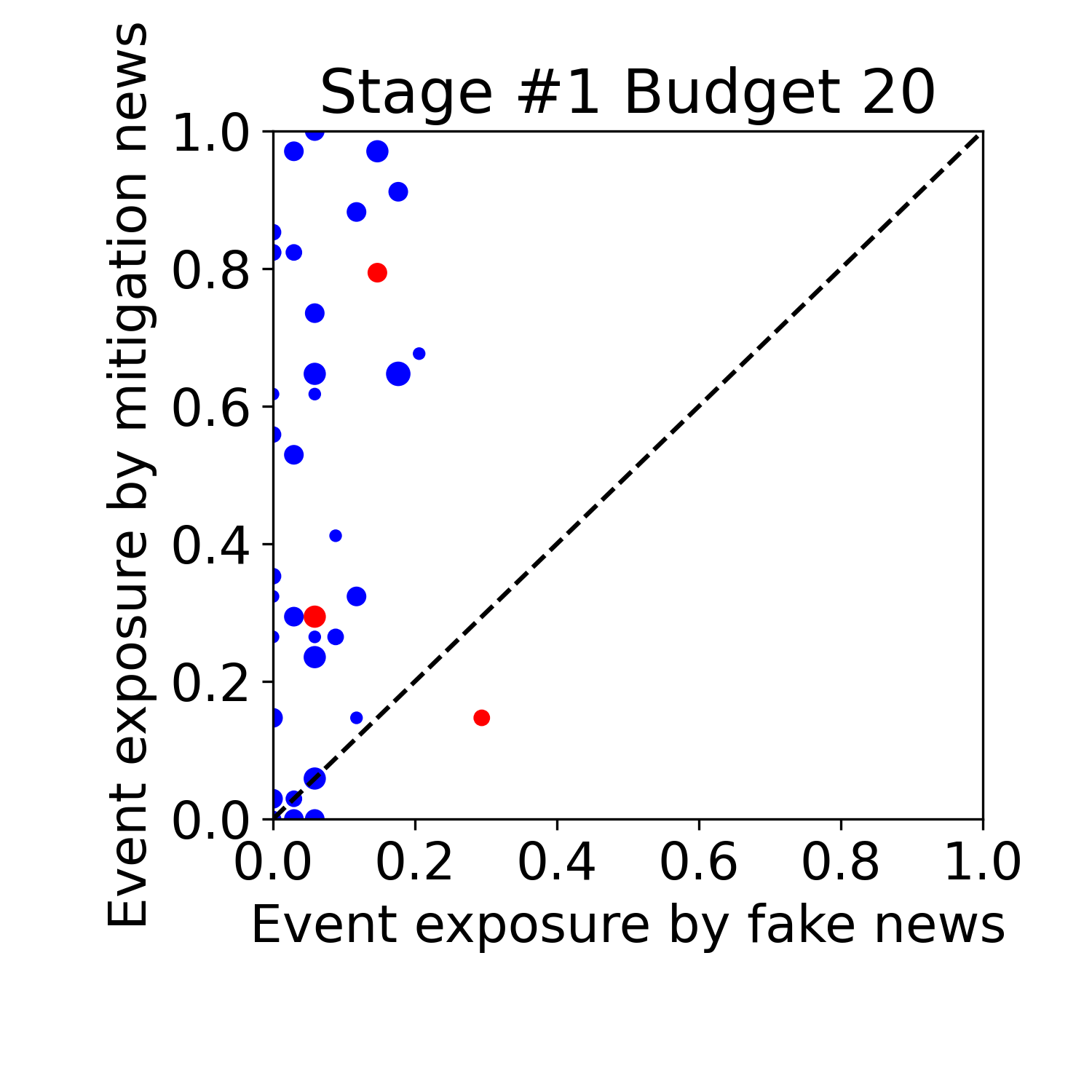} 
} 
\subfigure[Budget 20] { 
\label{fig:act5b20}     
\includegraphics[width=0.32\columnwidth]{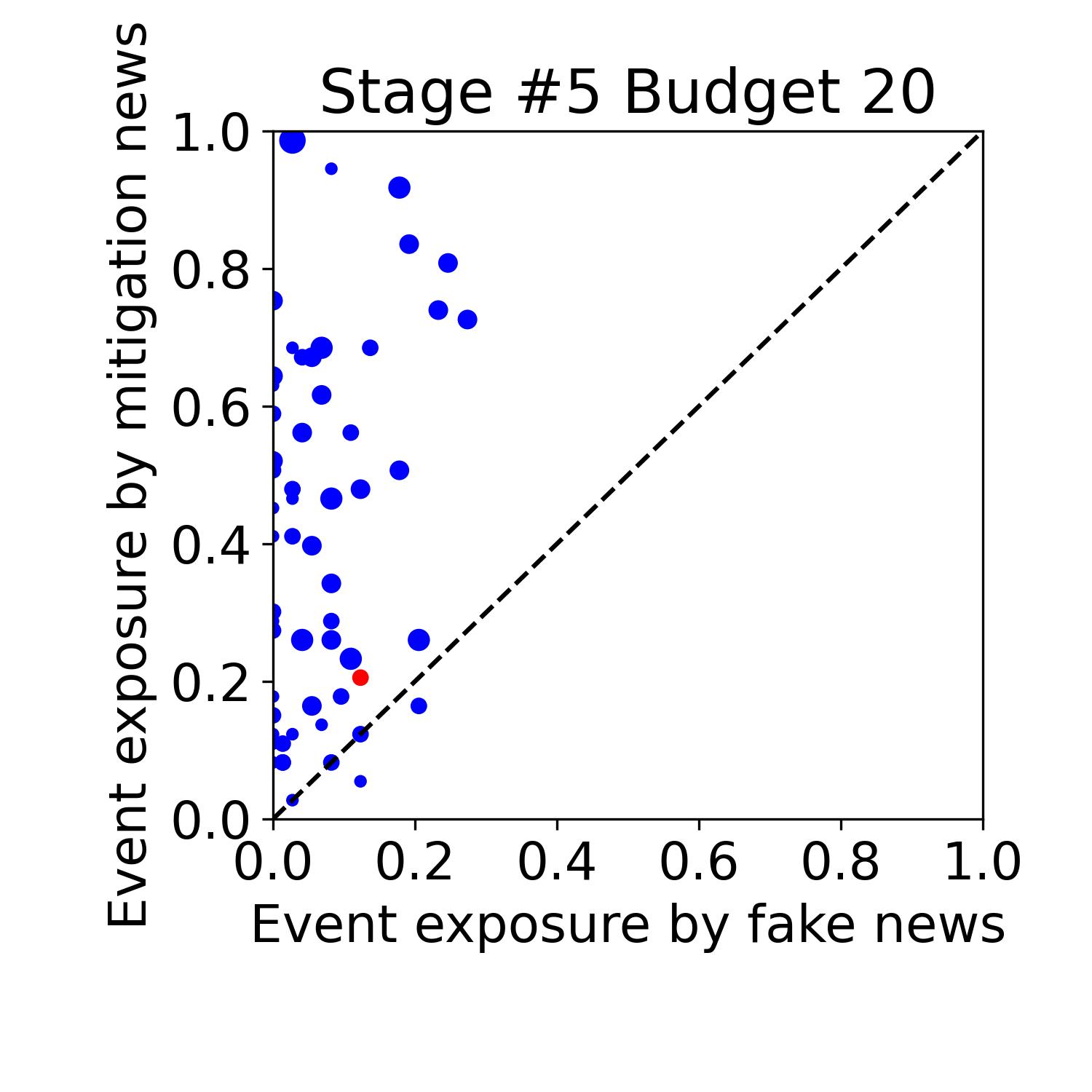}     
}    
\subfigure[Budget 20] { 
\label{fig:act9b20}     
\includegraphics[width=0.32\columnwidth]{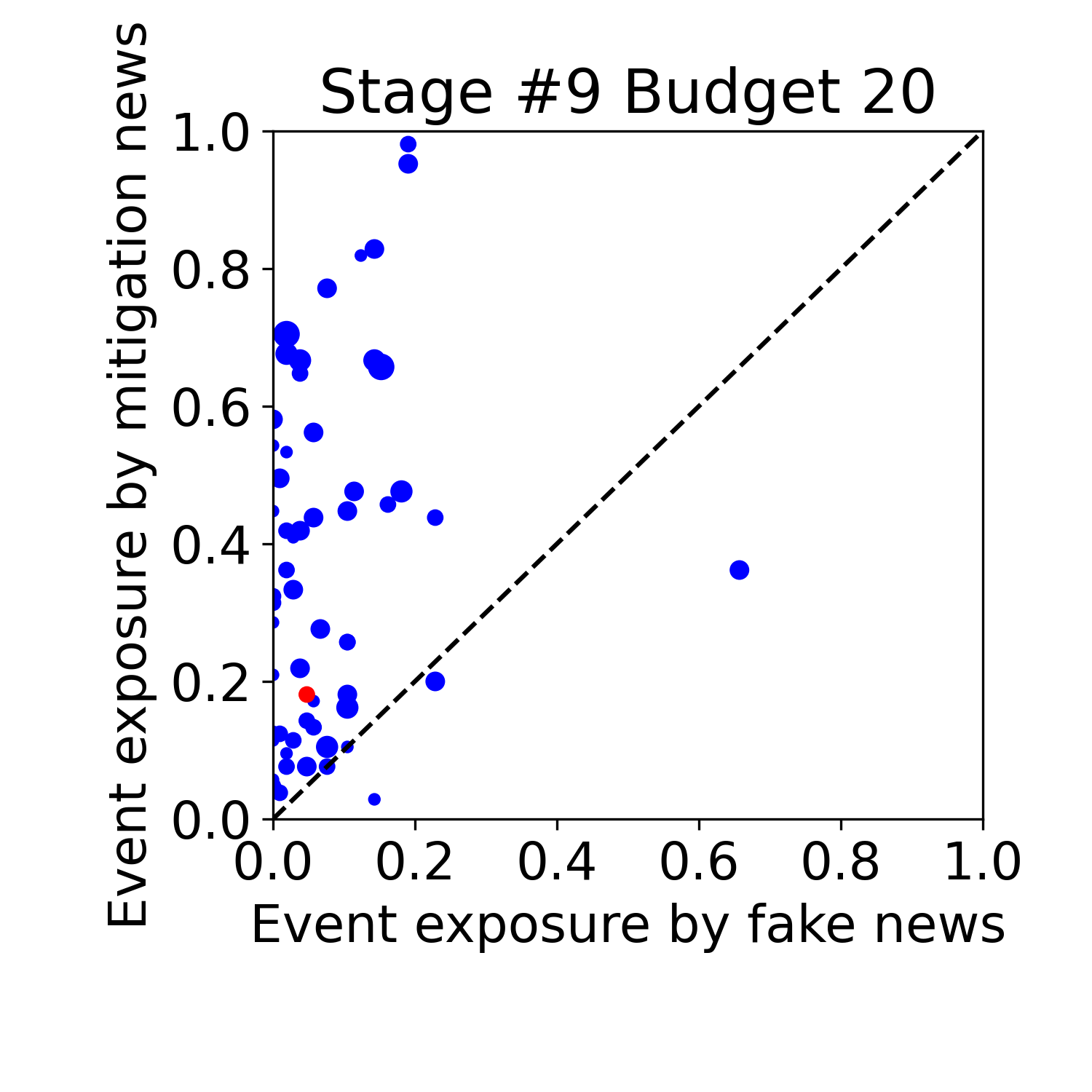}     
}
\vspace{-0.5cm}
\caption{Analysis of DQN-FSP at different budget settings.}
\vspace{-0.3cm}
    \label{fig:dumpact}
\end{figure*}

\begin{figure}
    \centering
\subfigure[FSP training and impact] {
 \label{fig:act1}     
\includegraphics[width=0.47\columnwidth]{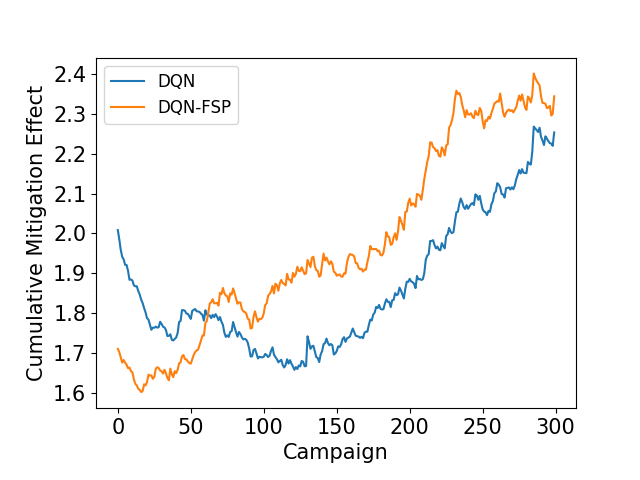} 
} 
\subfigure[Selecting debunkers] { 
\label{fig:act5}     
\includegraphics[width=0.47\columnwidth]{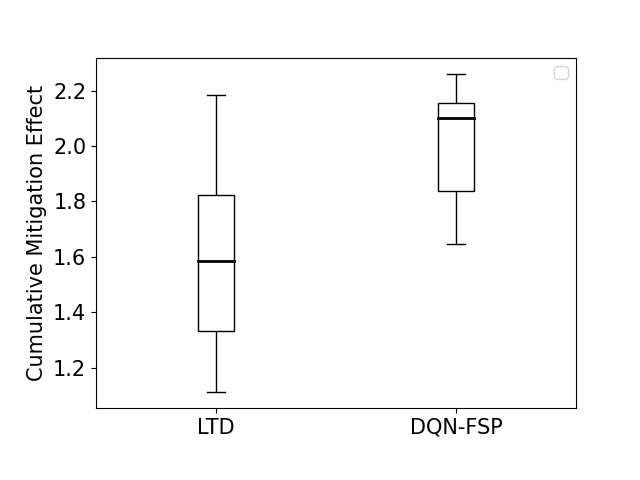}     
}    
\vspace{-0.5cm}
\caption{Analysis of DQN-FSP.}
\label{fig:convergence}
\vspace{-0.3cm}
\end{figure}



\subsubsection{Experiment Results} 
The performance is measured by the average cumulative mitigation effect using our DQN-FSP and baselines (3 runs on 100 mitigation campaigns). With the performance of \emph{RND} as the basis, Figure \ref{fig:real-param} shows the relative performance against \emph{RND} at different settings on different datasets. Wherever it is required, the relevant parameters are randomly set in the given ranges as discussed in Section \ref{sec:ses}.



Figure \ref{fig:real-param} (a)(d)(g) show the performance with respect to the density of the social network. It illustrates that our DQN-FSP outperforms baselines at almost all settings on different datasets. The performance trends in (d) and (g) are similar and are different from those in (a). The reason is that at different density, if the ratio between edge number and node number is 1 for GUR, the ratio is about 3 for PRI and PUT. Compared with Figure \ref{fig:syn-param} (a) where synthetic data has a ratio similar to that of GUR, we can observe it has a similar performance trend as GUR shown in Figure \ref{fig:real-param} (a). Moreover, comparing \emph{MAX-INF} with \emph{RND}, the performance of \emph{MAX-INF} declines when the number of edges per node increases. The reason is that all nodes have more influence and thus the influence of randomly selected nodes are comparable with that of nodes selected using \emph{MAX-INF}.   

Figure \ref{fig:real-param} (b)(e)(h) show the performance with respect to the average stage length. As mentioned in the synthetic experiment results, the longer stage length means the actions at the beginning of the stage will last longer and thus have more effect on the environment. The experiment results show our DQN-FSP outperforms baselines at different settings on different datasets.

Figure \ref{fig:real-param} (c)(f)(i) show the performance with respect to the number of stages. As discussed above, with more stages the agent has more opportunities to optimize mitigation and more chances to increase the intensity for selected debunkers. At all settings on all three topics, our DQN-FSP outperforms the baselines.



\vspace{-0.1cm}
\subsection{Analysis of DQN-FSP}\label{subsec:casestudy}
Various aspects of DQN-FSP are analysed next.
All discussions are based on results for the real-world data GUR. 

\subsubsection{Budget size} 
Intuitively, the performance of DQN-FSP is impacted by the budget at each stage. In Figure \ref{fig:dumpact}, $x$-axis represents the exposure of fake news and $y$-axis represents the exposure of true news where each point represents a user and the red points represent the source of fake news. At the end of a mitigation stage, ideally all users are above the diagonal, i.e., they receive more true news than fake news. To make it comparable, a mitigation campaign has 10 stages where the $k$-th stage starts at time $w_k$ and ends at time $w_{k+1}$.  Figure \ref{fig:dumpact} (a)(b)(c) illustrate the distribution of users based on the number of fake news and true news received at $w_2$, $w_6$ and $w_{10}$ (i.e. at the end of stages 1, 5 and 9),
when no mitigation is applied and users send and receive fake news and true news based on background intensity.
Clearly, many users are under the diagonal, that is, receiving more fake news than true news. Figure \ref{fig:dumpact} (d)(e)(f) present the distribution of users when mitigation campaigns are applied within budget 5 at the end of stages 1, 5 and 9. Clearly, more users are above the diagonal after stages 1, 5 and 9, but there still exist some users below the diagonal. Note that the total number of users is 98 (including 5 fake news spreaders) and the average user mitigation cost is 2.4. The budget 5 implies that debunkers are 2.13\% of all users. 

Comparing Figure \ref{fig:dumpact} (g)(h)(i) with Figure \ref{fig:dumpact} (d)(e)(f), the only difference is that budget at each stage is 10 and we observe that most users are above the diagonal after stage 1, and more users are moved above the diagonal after stages 5 and 9. Comparing Figure \ref{fig:dumpact} (j)(k)(l) with Figure \ref{fig:dumpact} (g)(h)(i), the only difference is that budget at each stage is increased to 20. It shows almost all users are above the diagonal. The performance between budget 10 and 20 is trivial. It indicates that budget 10 is sufficient. The results in Figure \ref{fig:dumpact} verify that, with the DQN-FSP mitigation policy, users exposed to more fake news will receive more true news and more budget results in more effective mitigation at an early stage. 





\subsubsection{Future state prediction training and impact} 
As discussed above, to minimize the mitigation overlap, we propose an RNN model to predict the future state that the currently selected debunkers may lead to. So, when the agent selects the next debunker, it will avoid those debunker candidates with overlapping mitigation effect. The future state prediction (FSP) accuracy plays a significant role in DQN-FSP. This experiment compared our DQN-FSP against baseline \emph{DQN} at different sizes of training data. From \ref{fig:convergence} (a), we observe that baseline \emph{DQN} has better performance than DQN-FSP when the training dataset has less than 50 campaigns (episodes). This is because the RNN model is not well trained yet and FSP accuracy is not sufficiently good. Once the training dataset has more than 50 campaigns, FSP has a better performance and in turn, the performance of DQN-FSP becomes better than \emph{DQN} consistently.

\subsubsection{Selecting debunkers} 
We compare our DQN-FSP with baseline \emph{LTD}~\cite{farajtabar2017fake} where the same set of debunkers are applied at different stages throughout a mitigation campaign. We have run \emph{LTD} for 50 times (i.e., 50 campaigns) where each campaign randomly selects users as debunkers within the same campaign budget (equally split to stages). The distribution of cumulative mitigation effects for the 50 runs is shown in  \ref{fig:convergence} (b). For DQN-FSP, we have trained 5 different mitigation policies since various settings are randomly selected in specified ranges. Using each of the trained mitigation policies, a mitigation campaign is executed within the same campaign budget as that for \emph{LTD} (but randomly split into stages). The distribution of cumulative mitigation effects for the 5 runs is shown in Figure \ref{fig:convergence} (b) as well. Clearly, the performance of DQN-FSP is significantly better than that of \emph{LTD}, which shows the benefit of selecting debunkers compared to the fixed debunker approach in previous studies~\cite{farajtabar2017fake} for the multi-stage campaign.








\subsection{Limitations}
Note that similar to existing studies (e.g., \cite{farajtabar2017fake}), our proposed mitigation policy assumes that the truth value -- true or fake -- for social media news posts are established and fed to the mitigation process. Errors of upstream fake news detection models therefore can propagate into the mitigation model. 
Future work can address this limitation in an end-to-end framework. 

\section{Conclusion}\label{conclusion}
This paper studied the problem of selecting debunkers for multi-stage fake news mitigation campaigns on social networks. 
We proposed a reinforcement learning framework to learn a mitigation policy that selects multiple debunkers dynamically within budget for each stage 
so that the selected debunkers can maximize the overall cumulative mitigation effect across stages. 
To address the issue of selecting debunkers from an exponentially large search space, we proposed a greedy algorithm 
with future state prediction so that debunkers are selected in a way that minimizes mitigation overlap and maximizes the overall mitigation effect. 
For future work, we will model other aspects of fake news propagation for more effective mitigation. 


\section{Acknowledgments}
This research was supported partially by the Australian Government through the Australian Research Council's Discovery Projects funding scheme (project: DP200101441, DP210100743) and Linkage Projects funding scheme (project: LP180100750).

\bibliographystyle{ACM-Reference-Format}
\balance
\bibliography{references}

\end{document}